\documentclass[12pt]{article}
\usepackage{epsfig}
\usepackage{amssymb,amsmath}
\usepackage{setspace}
\pdfoutput=1
\usepackage{subfigure}
\usepackage{amssymb,amsmath}
\usepackage{graphicx}
\usepackage{color}
\usepackage{cancel}
\usepackage[colorlinks=true
,urlcolor=blue
,citecolor=blue
,linkcolor=blue
,pagecolor=blue
,linktocpage=true
,pdfproducer=medialab
]{hyperref}
\def\bi{\begin{itemize}}
\def\ei{\end{itemize}}

\def\tl{\tilde l}

\def\alt{\lesssim}
\def\agt{\gtrsim}

\def\tl{\tilde l}

\def\tl{\tilde}

\usepackage{amsmath}

\newcommand{\bea}{\begin{eqnarray}}
\newcommand{\eea}{\end{eqnarray}}

\newcommand{\beq}{\begin{equation}}
\newcommand{\eeq}{\end{equation}}

 \makeatletter
\def\alt{\mathrel{\mathpalette\gl@align<}}
\def\agt{\mathrel{\mathpalette\gl@align>}}
\def\gl@align#1#2{\lower.6ex\vbox{\baselineskip\z@skip\lineskip\z@
\ialign{$\m@th#1\hfil##\hfil$\crcr#2\crcr\sim\crcr}}} \makeatother

\usepackage{longtable}

\makeatletter

\newcommand{\Rmnum}[1]{\expandafter\@slowromancap\romannumeral #1@}
\makeatother
\begin{document}
%

\begin{center}
\baselineskip 20pt {\Large\bf
Constraining Natural SUSY via the Higgs Coupling and the Muon Anomalous Magnetic Moment Measurements
}
\vspace{1cm}

{\large
Tianjun Li$^{a,b}$\footnote{E-mail: tli@itp.ac.cn},
Shabbar Raza$^{a}$\footnote{E-mail: shabbar@itp.ac.cn}, 
Kechen Wang$^{c}$\footnote{E-mail: kechen@ihep.ac.cn}
} \vspace{.5cm}

{\baselineskip 20pt \it $^a$
State Key Laboratory of Theoretical Physics and Kavli Institute for Theoretical Physics China (KITPC),
Institute of Theoretical Physics, Chinese Academy of Sciences,\\
Beijing, 100190, China \\
}
{\it $^b$
School of Physical Electronics, University of Electronic Science and Technology of China, Chengdu, 610054, China \\
}
{\it $^c$
Center for Future High Energy Physics, Institute of High Energy Physics,\\
Chinese Academy of Sciences, Beijing, 100049, China \\
}

\vspace{.5cm}

\vspace{1.25cm} {\bf Abstract}
\end{center}
We use the Higgs coupling and the muon  anomalous magnetic moment measurements to constrain the parameter space of the natural supersymmetry
in the Generalized Minimal Supergravity (GmSUGRA) model. We scan the parameter space of the GmSUGRA model with small electroweak fine-tuning measure ($\Delta_{\rm EW} \leq 100$). The parameter space after applying various sparticle mass bounds, Higgs mass bounds, B-physics bounds, the muon magnetic moment constraint, and the Higgs coupling constraint from measurements at HL-LHC, ILC, and CEPC, is shown in the planes of various interesting model parameters and sparticle masses. Our study indicates that the Higgs coupling and muon  anomalous magnetic moment measurements can constrain the parameter space effectively. It is shown that $\Delta_{\rm EW}\sim$ 30, consistence with all constraints,
and having supersymmetric contributions to the muon  anomalous magnetic moment within 1$\sigma$ can be achieved. The precision of $k_b$ and $k_{\tau}$ measurements at CEPC can bound  $m_A$ to be above 1.2 TeV and 1.1 TeV respectively. The combination of the Higgs coupling measurement and muon anomalous magnetic moment measurement constrain  $\tilde{e}_R$ mass to be in the range from 0.6 TeV to 2 TeV. The range of both $\tilde{e}_L$ and $\tilde{\nu}_e$ masses is 0.4 TeV $\sim$ 1.2 TeV. In all cases, the $\tilde{\chi}_1^0$ mass needs to be small (mostly $\leq$ 400 GeV). The comparison of bounds in the $\tan\beta - m_A$ plane shows that the Higgs coupling measurement is complementary to the direct collider searches for heavy Higgs when constraining the natural SUSY. A few mass spectra in the typical region of parameter space after applying all constraints are shown as well.

\thispagestyle{empty}

\newpage

\addtocounter{page}{-1}

\baselineskip 18pt

\section{Introduction}

Supersymmetry (SUSY) is the most promising scenario for new physics beyond the Standard Model (SM). It not only provides
the unification of the SM gauge couplings, but also gives solution to the gauge hierarchy problem of the SM.
Under the assumption of $R$-parity conservation, the lightest supersymmetric particle (LSP), such as the lightest neutralino can be a good cold dark matter candidate. 
 
A SM-like Higgs has been discovered with mass around $m_{h}\sim$ 125 GeV \cite{ATLAS, CMS} which is a crowning 
achievement and it
completes the SM. Though $m_h\sim$ 125 GeV is little bit heavy, but it is still consistent with the prediction of the Minimal Supersymmetric Standard Model (MSSM) of 
$m_h \leq$ 135 GeV \cite{Martin:1997ns}. This somewhat heavy Higgs requires the multi-TeV top squarks with small mixing or TeV-scale
top squarks with large mixing. Moreover, we have strong constraints on the parameter space in the Supersymmetric SMs (SSMs) from the SUSY searches at the Large Hadron Collider (LHC).
For example, the gluino mass $m_{\tilde g}$ should be heavier than about 1.7 TeV if the first
two-generation squark mass $m_{\tilde q}$ is around the gluino mass $m_{\tilde q} \sim m_{\tilde g}$,
and heavier than about 1.3 TeV for $m_{\tilde q} \gg m_{\tilde g}$~\cite{Aad:2014wea, Chatrchyan:2013wxa}.
The heavy SUSY spectrum and relatively heavy Higgs mass raise question about the naturalness of the MSSM. Some of the recent studies suggest that this problem
can be addressed and the naturalness of the MSSM is still there \cite{Drees:2015aeo,Ding:2015epa,Baer:2015rja,Batell:2015fma,AbdusSalam:2015uba,Barducci:2015ffa,Cohen:2015ala,Fan:2014axa,Leggett:2014hha,Dimopoulos:2014aua,Gogoladze:2013wva,Kribs:2013lua,Baer:2012cf,Gogoladze:2012yf,Baer:2012mv}. For instance, in Supernatural MSSM 
scenario~\cite{Du:2015una}, 
it was shown that no residual electroweak fine-tuning  (EWFT) left in the MSSM if we employ the No-Scale supergravity boundary 
conditions~\cite{Cremmer:1983bf} and Giudice-Masiero (GM) mechanism~\cite{Giudice:1988yz} even though one can have relatively heavy spectrum. But one of the major obstacle
for the above Supernatural SUSY studies is the $\mu$-term (higgsino mass parameter), which is generated by the GM mechanism
and then is proportional to the universal gaugino mass $M_{1/2}$.  The ratio $M_{1/2}/\mu$ is of order one but cannot be determined 
as an exact number. This problem was addressed in the M-theory inspired Next to MSSM (NMSSM) \cite{Li:2015dil}.
Another issue, related to the Higgs sector,  is the scrutiny of the properties of the SM-like Higgs boson predicted by the SM, such 
as its decay width, the Higgs couplings to the SM particles, and its spin and CP properties. This has already triggered new studies, experimental and
as well as theoretical~\cite{Dias:2015oza,Jez:2015wza,Cahill-Rowley:2013vfa}. Any deviations in the predicted properties of a SM-like Higgs boson may hint towards the physics Beyond the SM (BSM).
Moreover, besides the LHC, new $e^{+}e^{-}$ colliders have been proposed such as the International Linear Collider (ILC)  and 
Circular Electron Positron Collider (CEPC) where these Higgs properties can be studied with high precisions. Apart from looking
for physics at high energy colliders, one can also get glimpses of the BSM physics by using low energy precision measurements such as the 
measurements of the muon magnetic moment $(g-2)_{\mu}$. To address the $(g-2)_{\mu}$ anomaly between experiment 
and theory, new direct measurements of the muon magnetic moment with fourfold improvement
in accuracy have been proposed at Fermilab by the E989 experiment as well as Japan Proton Accelerator Research
Complex~\cite{Gray:2015qna,Venanzoni:2014ixa}. First results from E989 are expected around 2017/2018. These measurements will firmly establish or
constrain the new physics effects. Spurred by these developments, new studies have been done in order to explore this
opportunity~(For some latest studies, see Ref.~\cite{Gogoladze:2015jua,Chowdhury:2015rja,Harigaya:2015jba,Wang:2015rli,Kowalska:2015zja,Harigaya:2015kfa,Ajaib:2015ika,Athron:2015rva,Badziak:2014kea,Gogoladze:2014cha}).

In this paper, we try to study the parameter space of General Minimal Supergravity (GmSUGRA)~\cite{Li:2010xr,Balazs:2010ha} by imposing
 naturalness, Higgs coupling precision measurement and the muon $(g-2)$ measurement as constraints. Besides these constraints, 
we also demand that the parameter space is consistent with Higgs mass bounds, SUSY particle mass bounds and B physics constraints.
By concerning the naturalness of GmSUGRA, we will be probing the parameter space with low $\mu$ values. In this scenario one can expect to
have light higgsinos as the LSPs. But bino, wino or mixed DM may also be possible in some regions of parameter space. In addition to it,
one can  constrain the stop quark mass ranges~\cite{Brust:2011tb}. One can also probe the BSM physics by studying the Higgs couplings such as $hbb,h\tau\tau,htt,hWW,hZZ$ as functions of pseudo-scalar mass $m_A$. We will show that these precision measurements can constrain $m_A$ effectively. Stop quark masses can 
also be constrained by $hgg$ coupling while $h\gamma\gamma$ can constrain not only stop quark but also chargino masses.
On the other hand, it is a well-known fact that if SUSY provides solution to the muon $(g-2)_{\mu}$ discrepancy, sleptons and elctroweakinos
(charginos, bino, wino, and/or higgsinos) should be light \cite{Endo:2013bba}. In this study we see that if some parameters, such as $m_A$, cannot be constrained by the naturalness constraint, they can
be constrained by Higgs coupling precision measurements. Moreover, some parameters, such as stops and electroweakions, can be restricted by more than one
constraints. We hope that the ongoing and future experiments will be able to probe the BSM physics
and shed light on new avenues of physics.

In this paper, we restrict our solutions to $\Delta_{EW} \leq$ 100  which is a measure of Electroweak Fine-Tuning (EWFT) and will be
discussed later. We find that
the minimal value of  $\Delta_{EW}$ for a point satisfying  Higgs mass bounds, SUSY particle mass bounds and B-physics bounds (which we call
basic constraints) is about
8 with $\mu\sim$ 0.1 TeV, but it jumps to 20 after the application of the $(g-2)_{\mu}$ bounds and Higgs coupling precision measurement
bounds with $\mu\sim$ 0.140 TeV. The minimal light stop quark mass consistent with all the constraints is found to be around 0.7 TeV.
The pseudo-scalar mass $m_A$ can be constrained by using the $hbb$ ($h\tau\tau$) Higgs coupling precision measurements at the 
LHC-HL, ILC and CEPC in the mass bounds of, 0.4 (0.5) TeV, 1.1 (0.9) TeV and 1.2 (1.1) TeV, respectively. On the other hand, $hgg$ coupling
can constrain the light stop quark mass upto 0.5 TeV but by combining with $(g-2)_{\mu}$ constraint, it rises to 0.7 TeV as 
mentioned above. The deviations in $hWW$ and $hZZ$ are very small and beyond the sensitivity of the collider measurements. We also notice in our present scans, solutions that satisfy the basic constraints and the CEPC Higgs coupling constraint can have 
$\Delta_{EW}\sim$ 30-40 and contributions to $(g-2)_{\mu}$ within 1$\sigma$ of $\Delta a_{\mu}$ measurement. 
Slepton and electroweakino sectors are also constrained by the combination of constraints. For instance, $m_{\tilde e_{R}}$ and $m_{\tilde e_{L}}$ are
constrained to the mass ranges $[0.5,2]$ TeV and $[0.3,1.2]$ TeV, respectively. But to have contributions within 1$\sigma$ of $\Delta a_{\mu}$, we need
$m_{\tilde e_{R}}$ and $m_{\tilde e_{L}}$ in mass ranges $[0.8,1.3]$ TeV and $[0.4,0.5]$ TeV, receptively. For sneutrinos, the allowed mass ranges are
more or less in the same ranges as $m_{\tilde e_{L}}$. In electroweakino sector, the lightest neutralino, is confined in the mass range of 0.03 TeV to 0.5 TeV.
In our present scans, this ranges shrinks to even a smaller strip of 0.03 TeV to 0.3 TeV if we demand contributions within 1$\sigma$ of $\Delta a_{\mu}$. On the other hand, charginos can be as light as 0.1 TeV and the maximal allowed range is about 0.7 TeV. But for contributions 
better than 1$\sigma$ of $\Delta a_{\mu}$, we need chargino in the mass range $[0.16,0.22]$ TeV. Although we have not imposed relic density
constraint, but we do indicate regions of parameter space where the correct relic density can be achieved by the LSP neutralino annihilation and 
coannihilation mechanisms. For example, we show that there can be $A$ resonance and stau-neutralino coannihilation channels consistent with 
all the relevant constraints. We also note that the LSP neutralino can be higgsino, bino, wino or mixed DM. Furthermore, we indicate
the large mass gap in light stop and the LSP neutralino masses and comment on the 
possible detection of our solutions in the boosted stop scenario at the CEPC-SPPC \cite{Arkani-Hamed:2015vfh}. Finally, we display four benchmark points
as examples of our solutions.

The rest the of paper is organized as follows. In Section \ref{sec:theory}, we show the definition of the EWFT measure $\Delta_{\rm EW}$ and the theoretical expressions for the Higgs couplings and the muon anomalous magnetic moment in the GmSUGRA model. In Section \ref{sec:scan}, we give the phenomenological constraints and the scanning procedure. In Section \ref{sec:results}, we apply the constraints to the parameter space and discuss the numerical results. We conclude in Section \ref{sec:summary}.

\section{The GmSUGRA in the MSSM}
\label{sec:theory}
It was shown in ~\cite{Li:2010xr, Balazs:2010ha} that EWSUSY can be realized in the GmSUGRA model. In this scenario, the sleptons and charginos, bino, wino, and/or higgsinos are within one TeV while squarks and/or gluinos can be in several TeV mass 
ranges~\cite{Cheng:2012np}. In GmSUGRA, the GUT gauge group is $SU(5)$ and the Higgs field for the GUT symmetry breaking is in the
$SU(5)$  adjoint  representation~~\cite{Li:2010xr, Balazs:2010ha}. Since $\Phi$ can couple to the gauge field kinetic terms via 
high-dimensional operators, the gauge coupling relation and gaugino mass relation at the GUT scale will be modified after acquires
a Vacuum Expectation Value (VEV). The gauge
coupling relation and gaugino mass relation at the GUT scale are 
\begin{equation}
 \frac{1}{\alpha_2}-\frac{1}{\alpha_3} =
 k~\left(\frac{1}{\alpha_1} - \frac{1}{\alpha_3}\right)~,
\end{equation}
\begin{equation}
 \frac{M_2}{\alpha_2}-\frac{M_3}{\alpha_3} =
 k~\left(\frac{M_1}{\alpha_1} - \frac{M_3}{\alpha_3}\right)~,
\end{equation}
where $k$ is the index and equal to 5/3 in the simple GmSUGRA. We obtain a simple gaugino mass relation
\begin{equation}
 M_2-M_3 = \frac{5}{3}~(M_1-M_3)~,
\label{M3a}
\end{equation}
by assuming gauge coupling unification at the GUT scale ($\alpha_1=\alpha_2=\alpha_3$). 
The  universal gaugino mass relation $M_1 = M_2 = M_3$ in the mSUGRA, is just a special case of 
this general Eq.~\ref{M3a}. Choosing $M_1$ and $M_2$ to be free input parameters, which vary around 
several hundred GeV for the EWSUSY, we get $M_3$ from Eq.~(\ref{M3a})
\begin{eqnarray}
M_3=\frac{5}{2}~M_1-\frac{3}{2}~M_2~,
\label{M3}
\end{eqnarray}
which could be as large as several TeV or as small as several hundred GeV, depending
 on specific values of $M_1$ and $M_2$.
The general SSB scalar masses at the GUT scale are given 
in Ref.~\cite{Balazs:2010ha}. 
Taking the slepton masses as free parameters, we obtain the following squark masses 
in the $SU(5)$ model with an adjoint Higgs field
\begin{eqnarray}
m_{\tl{Q}_i}^2 &=& \frac{5}{6} (m_0^{U})^2 +  \frac{1}{6} m_{\tl{E}_i^c}^2~,\\
m_{\tl{U}_i^c}^2 &=& \frac{5}{3}(m_0^{U})^2 -\frac{2}{3} m_{\tl{E}_i^c}^2~,\\
m_{\tl{D}_i^c}^2 &=& \frac{5}{3}(m_0^{U})^2 -\frac{2}{3} m_{\tl{L}_i}^2~,
\label{squarks_masses}
\end{eqnarray}
where $m_{\tl Q}$, $m_{\tl U^c}$, $m_{\tl D^c}$, $m_{\tl L}$, and  $m_{\tl E^c}$ represent the scalar masses of
the left-handed squark doublets, right-handed up-type squarks, right-handed down-type squarks,
left-handed sleptons, and right-handed sleptons, respectively, while $m_0^U$ is the universal  
scalar mass, as in the mSUGRA. In the Electroweak SUSY (EWSUSY), $m_{\tl L}$ and $m_{\tl E^c}$ are both within 1 TeV, resulting in 
light sleptons. Especially, in the limit $m_0^U \gg m_{\tl L/\tl E^c}$, we have the approximated 
relations for squark masses: $2 m_{\tl Q}^2 \sim m_{\tl U^c}^2 \sim m_{\tl D^c}^2$. In addition, 
the Higgs soft masses $m_{\tl H_u}$ and $m_{\tl H_d}$, and the  trilinear soft terms
 $A_U$, $A_D$ and $A_E$ can all be free parameters from the GmSUGRA~\cite{Balazs:2010ha, Cheng:2012np}.
\subsection{The Electroweak Fine-Tuning}
\label{ewft}

GmSUGRA model offers solution to the Electroweak Fine-Tuning (EWFT) problem ~\cite{Li:2014dna}. We use  ISAJET 7.85 \cite{ISAJET} 
to calculate the fine-tuning conditions at the EW scale $M_{\rm EW}$. The $Z$ boson mass $M_Z$, after including the one-loop effective potential contributions to the tree-level MSSM Higgs potential, is given by

\begin{equation}\label{eqn:Finetunning}
\frac{m_Z^2}{2} = \frac{m_{H_d}^2 + \Sigma_d^d - (m_{H_u}^2+\Sigma_u^u)~{\rm tan}^2\beta }{{\rm tan}^2\beta - 1} - \mu^2,
\end{equation}

\noindent where $\Sigma_u^u$ and $\Sigma_d^d$ denote the corrections to the scalar potential coming from the one-loop effective potential defined in \cite{Baer:2012mv} while  $m_{H_u}$ and $m_{H_d}$ are the Higgs soft masses. $\tan\beta \equiv \langle H_u \rangle / \langle H_d \rangle $ is the ratio of the Higgs VEVs. 
The largest contribution to $\Sigma_u^u$ comes from top squarks ($\tilde t_{1,2}$): 
$\Sigma_u^u \sim \frac{3y_{t}^2}{16\pi^2}\times m_{\tilde t_{1,2}}^2 log (m_{\tilde t_{1,2}/Q^{2}})$, where $y_t$ is the top Yukawa coupling and $Q=\sqrt{m_{\tilde t_1} m_{\tilde t_2}}$ \cite{Baer:2012up}. On the other hand, $m_{H_d}^2$ and $\Sigma_d^d$ terms are suppresed by $\tan^2\beta$. This allows one to have large $m_{H_d}^2$ and hence large $m_{A}^2$ values without agrivate fine-tuning problem~\cite{Baer:2012up}. We discuss it
more in the later part of the paper. All the parameters in Eq.~(\ref{eqn:Finetunning}) are defined at the
electroweak scale $M_{\rm EW}$. In order to measure the EWFT condition we follow \cite{Baer:2012mv} and use the definitions
\begin{equation}\label{eqn:Chmu}
C_{H_d} \equiv |m_{H_d}^2/({\rm tan}^2 \beta -1)|, C_{H_u} \equiv |-m_{H_u}^2 ({\rm tan}^2 -1)|, C_{\mu} \equiv |-{\mu}^2|,
\end{equation}
with each $C_{\Sigma_{u,d}^{u,d}(k)}$ less than some characteristic value of order $M_Z^2$. Here, $k$ labels the SM and SUSY particles that contribute to the one-loop Higgs potential. For the fine-tuning measure, we define

\begin{equation}\label{eqn:FTew}
\Delta_{\rm EW} \equiv {\rm max} (C_k)/(M_Z^2/2).
\end{equation}

Note that $\Delta_{\rm EW}$ only depends on the weak-scale parameters of the SUSY models, and then is fixed by the particle spectra. Hence, it is independent of how the SUSY particle masses arise. The lower values of $\Delta_{\rm EW}$ corresponds to less fine tuning, for example, $\Delta_{\rm EW} = 10$ implies $\Delta_{\rm EW}^{-1} = 10\%$ fine tuning. In addition to $\Delta_{\rm EW}$, ISAJET also calculates
$\Delta_{\rm HS}$, which is a measure of fine-tuning at the High Scale (HS) like the GUT scale in our case \cite{Baer:2012mv}. The HS fine tuning measure $\Delta_{\rm HS}$ is given as follows
\begin{equation}\label{eqn:FThs}
\Delta_{\rm HS} \equiv {\rm max} (B_i)/(M_Z^2/2).
\end{equation}
For definition of $B_i$ and more details, see Ref.~\cite{Baer:2012mv}.

\subsection{Higgs Couplings}
\label{subsec:HiggsCoupling}

In this subsection, we show the theoretical expressions of the Higgs couplings in the GmSUGRA model. Their deviations from the SM Higgs couplings are parametrized by the ratio $k_i \equiv g_{hii}^{\rm SUSY} / g_{hii}^{\rm SM}$, where $i=W,Z,b,\tau,t,g, \gamma$. Here $g_{hii}^{\rm SM}$ is the SM Higgs couplings, while $g_{hii}^{\rm SUSY}$ is the SUSY Higgs couplings.

In general, the Higgs couplings to W and Z gauge bosons mainly depend on the angle $\beta$ and the mixing angle $\alpha$ between the SM-like Higgs boson and the heavier CP-even Higgs boson. From the reference \cite{Carena:2001bg}, in the decoupling limit of $m_A \gg m_Z$, we can have the relation
\begin{eqnarray}
\cos(\alpha+\beta) \sim {\cal O}\left( \frac{m_Z^2\sin4\beta}{2m_A^2} \right)
\sim {\cal O}\left(- \frac{m_Z^2}{m_A^2} \frac{2}{\tan \beta} \right).
\label{eq:cosaplb}
\end{eqnarray}
Here we have terminated the above expression upto the order ${\cal O}\left( \frac{m_Z^2}{m_A^2} \right)$ and used the identity $\sin 4\beta \approx - \frac{4}{\tan \beta}$ for large $\tan\beta$.

Since $\cos(\alpha+\beta) \approx 0$ when $m_A \gg m_Z$, we then have
\begin{equation}
\sin(\alpha+\beta) \approx 1-\frac12\cos^2(\alpha+\beta) \sim 1-{\cal O}\left(\frac{m_Z^4}{m_A^4} \frac{2}{\tan^2 \beta} \right).
\label{eq:sinaplb}
\end{equation}

Following the reference \cite{Bae:2015nva}, in the MSSM the Higgs couplings to W and Z gauge bosons are given as $g_{hVV}^{\rm SUSY}=g_{hVV}^{\rm SM} \sin(\alpha+\beta)$, where V = W, Z.
Therefore, these deviations can be expressed as
\begin{equation}
k_V \equiv \frac{g_{hVV}^{\rm SUSY}}{g_{hVV}^{\rm SM}}  = \sin(\alpha+\beta) \sim 1-{\cal
O}\left(\frac{m_Z^4}{m_A^4} \frac{2}{\tan^2 \beta} \right),\, \rm{for}\,\, V = W, Z.
\label{eqn:kv}
\end{equation}

The deviations of the Higgs couplings to fermions ($b, \tau, t$) are given in \cite{Bae:2015nva} as

\begin{eqnarray}
k_b&=&\sin(\alpha+\beta)-\frac{\cos(\alpha+\beta)}{1+\Delta_b}\left\{
\tan\beta -\Delta_b \cot \beta +(\tan\beta + \cot\beta )\frac{\delta f_b}{f_b} \right\} \nonumber \\ 
k_{\tau}&=&\sin(\alpha+\beta)-\frac{\cos(\alpha+\beta)}{1+\Delta_{\tau}}\left\{
\tan\beta -\Delta_{\tau} \cot \beta +(\tan\beta + \cot\beta )\frac{\delta f_{\tau}}{f_{\tau}} \right\} \nonumber \\ 
k_t&=&\sin(\alpha+\beta)+\frac{\cos(\alpha+\beta)}{1+\Delta_t} 
\left\{ (1+\Delta_t)\cot \beta -(1+\cot^2\beta)\frac{\Delta f_t}{f_t} \right\}
\label{eqn:kb}
\end{eqnarray}

Plugging in the above expressions of $\sin(\alpha+\beta)$ and $\cos(\alpha+\beta)$ and terminating the expressions upto the order ${\cal O}\left( \frac{m_Z^2}{m_A^2} \right)$, finally we can get
\begin{eqnarray}
k_{b,\tau} &\sim& 1-{\cal O}\left(\frac{m_Z^4}{m_A^4} \frac{2}{\tan^2 \beta} \right) + {\cal O} \left( \frac{m_Z^2}{m_A^2} \frac{2}{\tan \beta} \tan\beta \right) \sim 1 + {\cal O} \left( 2 \frac{m_Z^2}{m_A^2} \right) ~,
\label{eqn:kbktau}\\
k_t &\sim& 1-{\cal O}\left(\frac{m_Z^4}{m_A^4} \frac{2}{\tan^2 \beta} \right) - {\cal O}\left( \frac{m_Z^2}{m_A^2} \frac{2}{\tan \beta} \cot\beta \right) \sim 1-{\cal O}\left(\frac{m_Z^2}{m_A^2} \frac{2}{\tan^2 \beta} \right)~.
\label{eqn:kt}
\end{eqnarray}

Furthermore, in the MSSM the deviation in the effective Higgs couplings to gluons are dominantly induced by the stop loop contribution, which can be
approximately expressed as \cite{Arvanitaki:2011ck,Carmi:2012yp,Blum:2012ii} 
\begin{equation}
k_g \approx 1+\frac{m_t^2}{4}\left[
\frac{1}{m_{\tilde{t}_1}^2}+\frac{1}{m_{\tilde{t}_2}^2}
-\frac{X_{t}^2}{m_{\tilde{t}_1}^2m_{\tilde{t}_2}^2}
\right] ~,
\label{eqn:kg}
\end{equation}
where $X_t = \mid A_t-\mu/\tan\beta \mid$ is the stop mixing parameter.

The effective Higgs couplings to photons are much more complicated. In the SM, it is dominated by the $W$ boson loop contribution,
while in the MSSM, $k_\gamma$ gets contributions from all charged particles, including charged Higgs, stops and charginos. 

\subsection{The anomalous magnetic moment of the muon}
\label{subsec:g-2}

The theoretical value of the anomalous muon magnetic moment $a_{\mu}=(g-2)_{\mu}/2$ within the SM can be calculated to within sub-parts-per-million precision \cite{Davier:2010nc}. A comparison between the theoretical calculation and the experimental measurement of $a_{\mu}$ may reveal, though indirectly, traces for the physics beyond the SM. The discrepancy can be quantifized as follows~\cite{Bennett:2006fi}
\begin{equation}\label{eqn:ThDeltaAmu}
\Delta a_{\mu} \equiv a_{\mu}({\rm exp}) - a_{\mu}({\rm SM}) = (28.7 \pm 8.0) \times 10^{-10}~.~\,
\end{equation}

Moreover, using \cite{Hagiwara:2011af} for contributions of the hadronic vacuum polarization, and \cite{Prades:2009tw} for the hadronic light-by-light contribution, the discrepancy can be calculated as $\Delta a_{\mu} = (26.1 \pm 8.0) \times 10^{-10}$. Either way, $a_\mu$ has a $\sim 3 \sigma$ deviation from its SM value, providing a possible hint of new physics.

SUSY can address this discrepancy. At the EW scale, the main contributions to $\Delta a_{\mu}$ come from the neutralino-slepton and chargino-sneutrino loops and are given as
\begin{equation}\label{eqn:deltaAmuSUSY}
\Delta a_{\mu}^{\rm SUSY} \sim \frac{M_i\, \mu\, \rm{tan} \beta}{m_{\rm SUSY}^4 },
\end{equation}
where $M_i$ (i = 1, 2) are the weak scale gaugino masses, $\mu$ is the higgsino mass parameter, $\rm{tan}\beta \equiv \langle H_u \rangle / \langle H_d \rangle$, and $m_{\rm SUSY}$ is the sparticle mass circulating in the loop. For a review of the constraints on $\Delta a_{\mu}$ given by SUSY collider searches, see \cite{Padley:2015uma}.

\section{Phenomenological Constraints and Scanning Procedure}
\label{sec:scan}

We use the ISAJET 7.85 package \cite{ISAJET} to perform random scans.
The Higgs coupling ratios $k_i$ are also calculated by this package.
We scan over the parameter space given below
\begin{eqnarray}\label{eqn:scan}
0\, {\rm GeV} &\leq& m_0^U \leq 9000\, {\rm GeV}, \nonumber \\
100\, {\rm GeV} & \leq & M_1 \leq 2000\, {\rm GeV}, \nonumber \\
100\, {\rm GeV} & \leq & M_2 \leq 2100\, {\rm GeV}, \nonumber \\
100\, {\rm GeV} & \leq & m_{\tilde{L}} \leq 1200\, {\rm GeV}, \nonumber \\
100\, {\rm GeV} & \leq & m_{\tilde{E}^c} \leq 1200\, {\rm GeV}, \nonumber \\
100\, {\rm GeV} & \leq & \mu \leq 1500\, {\rm GeV}, \nonumber \\
0\, {\rm GeV} & \leq & m_{A} \leq 9500\, {\rm GeV}, \nonumber \\
-16000\, {\rm GeV} & \leq & A_U = A_D \leq 18000\, {\rm GeV}, \nonumber \\
-6000\, {\rm GeV} & \leq & A_E \leq 6000\, {\rm GeV}, \nonumber \\
2 & \leq & {\rm tan}\beta \leq 60.
\end{eqnarray}

When scanning the parameter space, we consider $\mu > 0$ and use $m_t =$ 173.3 GeV \cite{ATLAS:2014wva}. We use $m_b^{\bar{\rm DR}}(M_Z)$ = 2.83 GeV as it is hard-coded into ISAJET. We employ the Metropolis-Hastings algorithm as described in \cite{Belanger:2009ti} during our scanning. In rest of the paper, we will use the notations $A_t$, $A_b$, $A_\tau$ for $A_U$, $A_D$, and $A_E$, respectively.

After collecting the data, we apply the following constraints.

\noindent
\textbf{(\Rmnum{1}) Basic constraints:}

(a) The Radiative Electroweak Symmetry Breaking (REWSB).

(b) One of the neutralinos is the LSP.

(c) Sparticle masses.

We employ the LEP2 bounds on sparticle masses 
\begin{eqnarray}\label{eqn:spMassLEP2}
m_{\tilde{t}_1}, m_{\tilde{b}_1} & \geq & 100\, {\rm GeV}, \nonumber \\
m_{\tilde{\tau}_1} & \geq & 105\, {\rm GeV}, \nonumber \\
m_{\tilde{\chi}_1^{\pm}} & \geq & 103\, {\rm GeV}.
\end{eqnarray}

We also apply the following bounds from the LHC
\begin{eqnarray}\label{eqn:spMassLHC}
1.7\, {\rm TeV} \leq m_{\tilde{g}}\,\, ({\rm for}\,\, m_{\tilde{g}} \sim m_{\tilde{q}}) &\cite{Aad:2014wea, Chatrchyan:2013wxa}, \nonumber \\
1.3\, {\rm TeV} \leq m_{\tilde{g}}\,\, ({\rm for}\,\, m_{\tilde{g}} \ll m_{\tilde{q}}) &\cite{Aad:2014wea, Chatrchyan:2013wxa}, \nonumber \\
300\, {\rm GeV} \leq m_A &\cite{Bae:2015nva}.
\end{eqnarray}

(d) Higgs mass.

We use the following Higgs mass bound from the LHC
\begin{eqnarray}\label{eqn:higgsMassLHC}
123\, {\rm GeV} \leq m_h \leq 127\, {\rm GeV}\,  \cite{ATLAS, CMS}.
\end{eqnarray}

(e) B-physics.

We use the IsaTools package \cite{bsg, bmm} and implement the following B-physics constraints
\begin{eqnarray}\label{eqn:Bphysics}
1.6\times 10^{-9} \leq{\rm BR}(B_s \rightarrow \mu^+ \mu^-) 
  \leq 4.2 \times10^{-9} \;(2\sigma)~~&\cite{CMS:2014xfa} ~,~& 
\\ 
2.99 \times 10^{-4} \leq 
  {\rm BR}(b \rightarrow s \gamma) 
  \leq 3.87 \times 10^{-4} \; (2\sigma)~~&\cite{Amhis:2014hma}~,~ &  
\\
0.70\times 10^{-4} \leq {\rm BR}(B_u\rightarrow\tau \nu_{\tau})
        \leq 1.5 \times 10^{-4} \; (2\sigma)~~&\cite{Amhis:2014hma}~.~&
\end{eqnarray}

(f) Fine-tuning.

In this paper, since we  consider the natural SUSY, therefore, the following constraint for fine-tuning measure $\Delta_{\rm EW}$ is applied.
\begin{eqnarray}\label{eqn:deltaEW}
\Delta_{\rm EW} \leq 100.
\end{eqnarray}

\noindent
\textbf{(\Rmnum{2}) Muon anomalous magnetic moment constraint}

We also apply the following bounds for the muon anomalous  magnetic moment measurement
\begin{eqnarray}\label{eqn:deltaAmu}
4.7 \times 10^{-10} \leq \Delta a_{\mu} \leq 52.7 \times 10^{-10} \,\, (3\sigma) &\cite{Davier:2010nc}.
\end{eqnarray}

\noindent
\textbf{(\Rmnum{3}) Higgs coupling constraints}

The discovery of the SM-like Higgs boson with mass around 125 GeV provides the opportunity to extract the new physics indrectly by measuring the Higgs couplings (and other properties) precisely at a ``Higgs factory". For this study, we mainly consider two proposed future $e^+ e^-$ colliders which are able to produce a large number of Higgs events: the International Linear Collier (ILC) and the Circular Electron-Positron Collider (CEPC). As a linear $e^+ e^-$ collider, the ILC is designed to adopt the polarized beams technology and can reach a high center of mass energy $\sqrt{s} =$ 500 GeV \cite{Baer:2013cma}. The CEPC, however, so far is focusing on $\sqrt{s} =$ 240 GeV. Its proposed intergrated luminisity is $5\, {\rm ab}^{-1}$ over a running time of 10 years with 2 Interaction Points (IP) \cite{preCDR:Physics, preCDR:Accelerator}. Furthermore, the CEPC is designed to be upgraded to a 100 TeV Super Proton-Proton Collider (SPPC) finally.

Besides ILC and CEPC, we also consider the Higgs coupling measurements at the High-Luminosity LHC (HL-LHC) and the $e^+ e^-$ mode of the CERN Future Circular Collider which has 4 IP (FCC-ee (4 IP)). We use the precisions at CEPC (2 IP) from the Table 3.12 of \cite{preCDR:Physics}. The precisions at HL-LHC are given by the Table 3 of \cite{Atlas:2014susyCouplings}, while the precisions at the ILC and FCC-ee (4 IP) are given by the Table 1-16 of \cite{Dawson:2013bba}. We accumulate all these precisions of the Higgs coupling measurement in percentage at these colliders and list them in Table \ref{tab:HiggsCplP}. 
By applying these different sets of precisions, we can see the improvement in constraining the new physics with better precisions. It is worth noting that the precisions we listed in this table are mostly obtained by the 10-parameter fitting scheme. It is model-independent and the experimental observables are fit with 10 free parameters. One may get better precisions by using the more constraining fitting scheme with smaller number of free parameters (for example, the 7-parameter fitting scheme) or by combining the precisions of different colliders. This is beyond the scope of this study and we will not discuss it in this study.

\begin{table}[h]
\centering
\begin{tabular}{ccccc}
\hline
Collider & HL-LHC & ILC & CEPC (2 IP) & FCC-ee (4 IP)\\
$\sqrt{s}$ (GeV) & 14000 & 500 & 240 & 240 \\
$\mathcal{L}$ (${\rm fb}^{-1}$)& 3000 & 500 & 5000 & 10000 \\ 
polarization ($e^-$, $e^+$) & - & (-0.8, +0.3) & (0, 0) & (0, 0) \\ 
\hline
$k_g$ & 9.1 & 2.3 & 1.5 & 1.1 \\ 
$k_W$ & 5.1 & 1.2 & 1.2 & 0.85 \\ 
$k_Z$ & 4.4 & 1.0 & 0.26 & 0.16 \\ 
$k_{\gamma}$ & 4.9 & 8.4 & 4.7 & 1.7 \\ 
$k_b$ & 12 & 1.7 & 1.3 & 0.88 \\ 
$k_{\tau}$ & 9.7 & 2.4 & 1.4 & 0.94 \\ 
$k_t$ & 11 & 14 & - & - \\ \hline
\end{tabular}
\caption{Summary for the precisions of Higgs boson coupling measurements in percentage at different colliders.}
\label{tab:HiggsCplP}
\end{table}

\section{Numerical Results}
\label{sec:results}
In this section we present results of our scans.
\subsection{Naturalness}
\label{results_Naturalness}

In this subsection, we show the fine-tuning measure $\Delta_{\rm EW}$ as a function of those parameters which are related to this work. We apply various constraints discussed in Section~\ref{sec:scan} and restrict the points to $\Delta_{EW} \le$ 100.
\begin{figure}[htb!]
\centering
\subfigure{
\includegraphics[totalheight=5.5cm,width=7.cm]{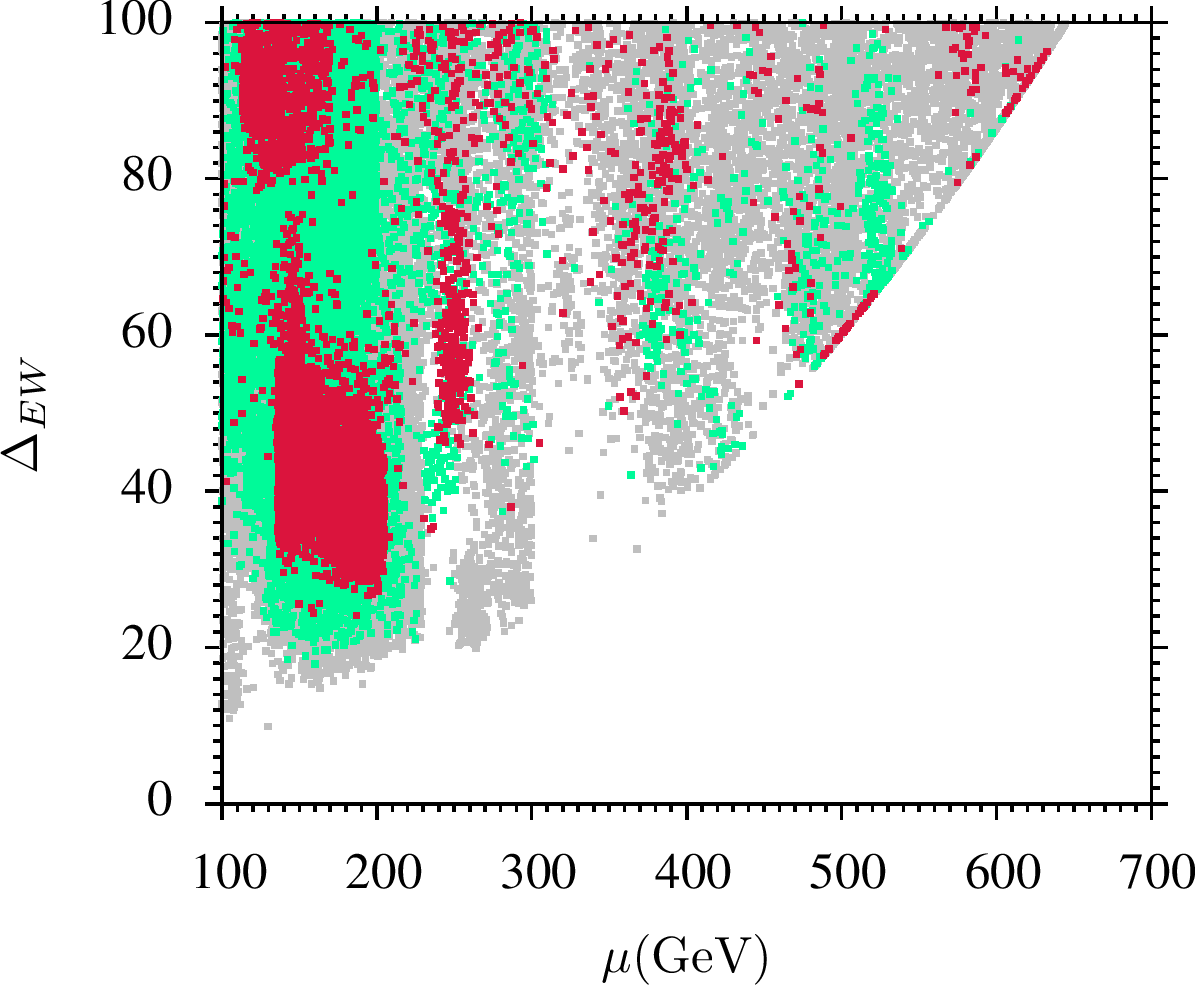}
}
\subfigure{
\includegraphics[totalheight=5.5cm,width=7.cm]{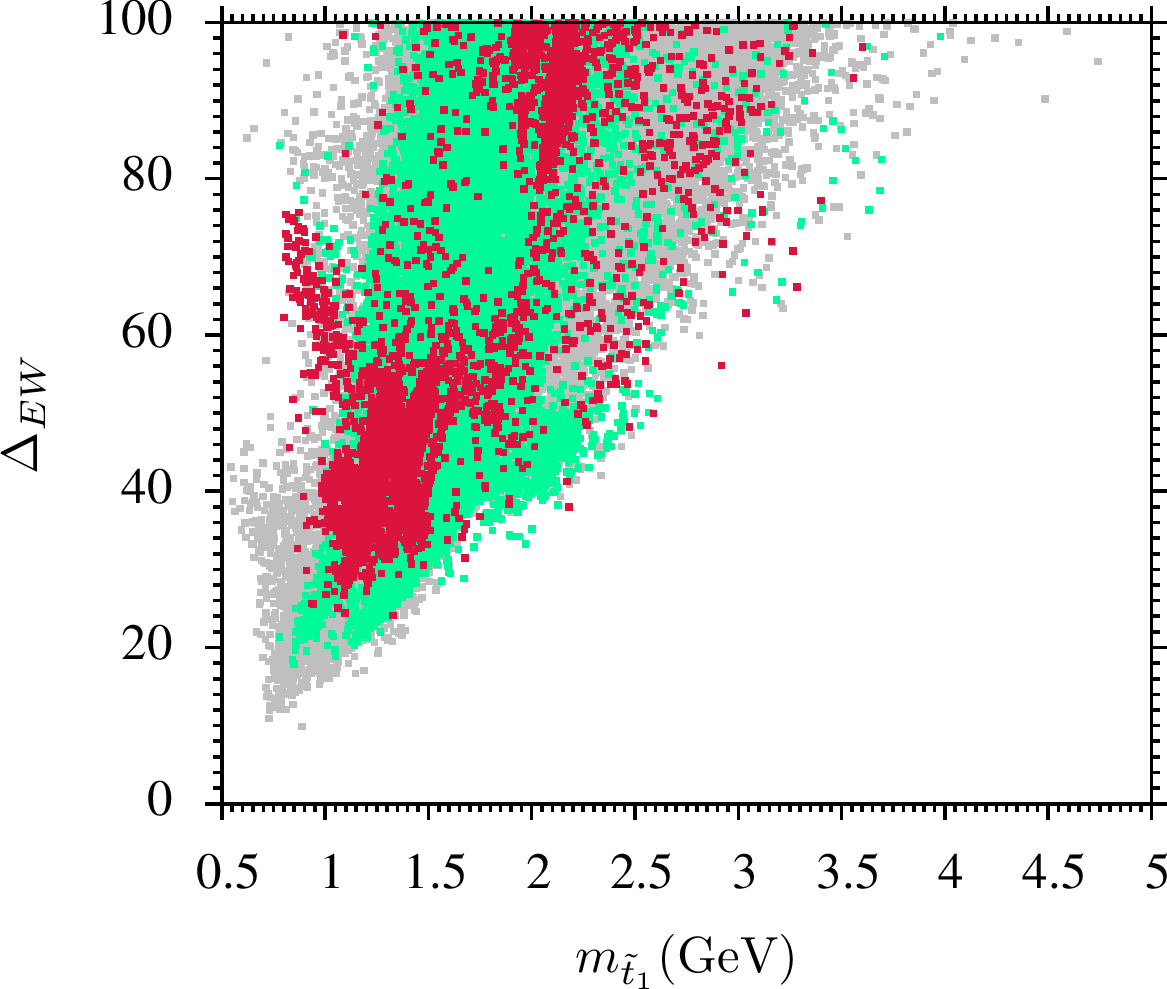}
}
\subfigure{
\includegraphics[totalheight=5.5cm,width=7.cm]{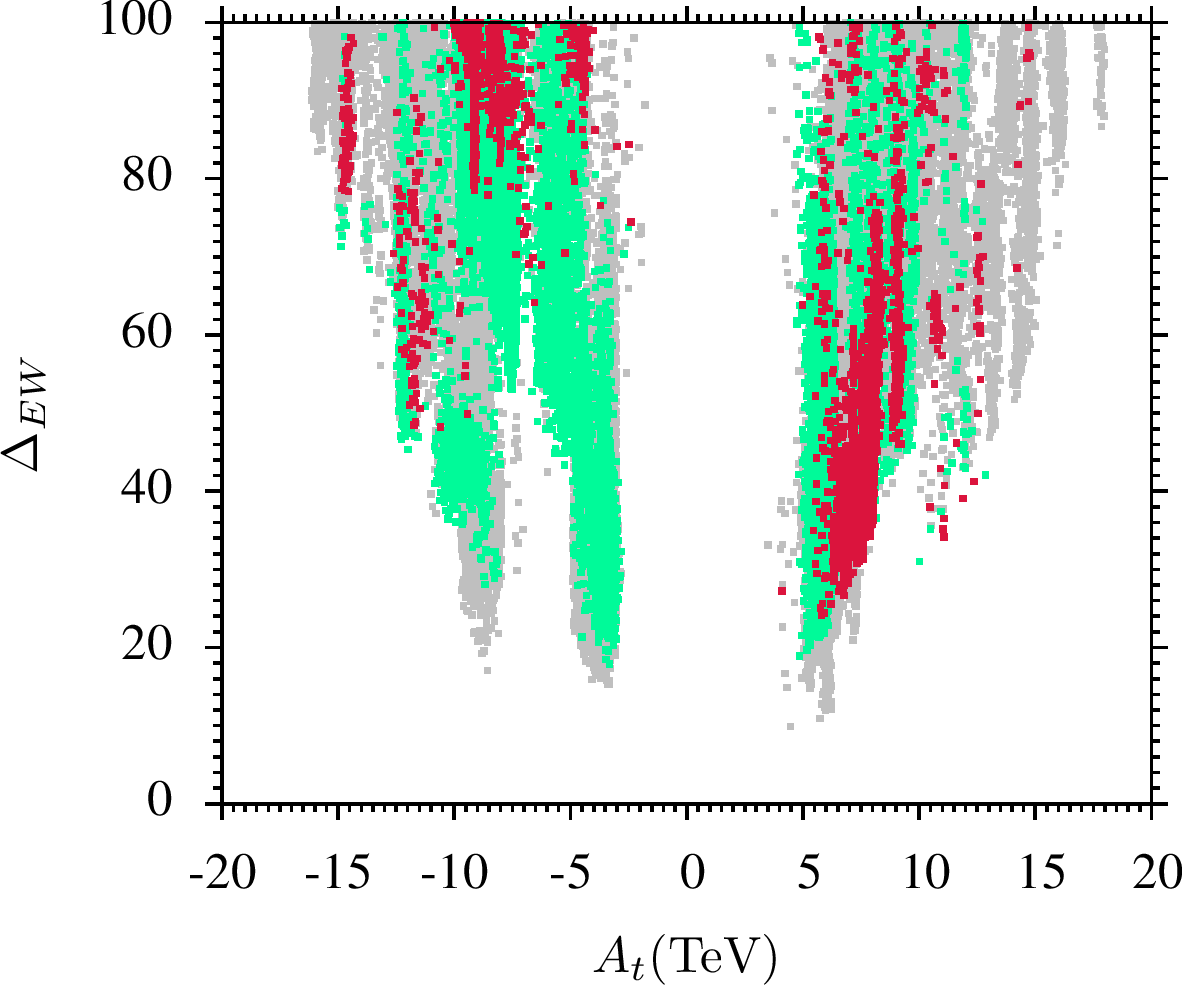}
}
\subfigure{
\includegraphics[totalheight=5.5cm,width=7.cm]{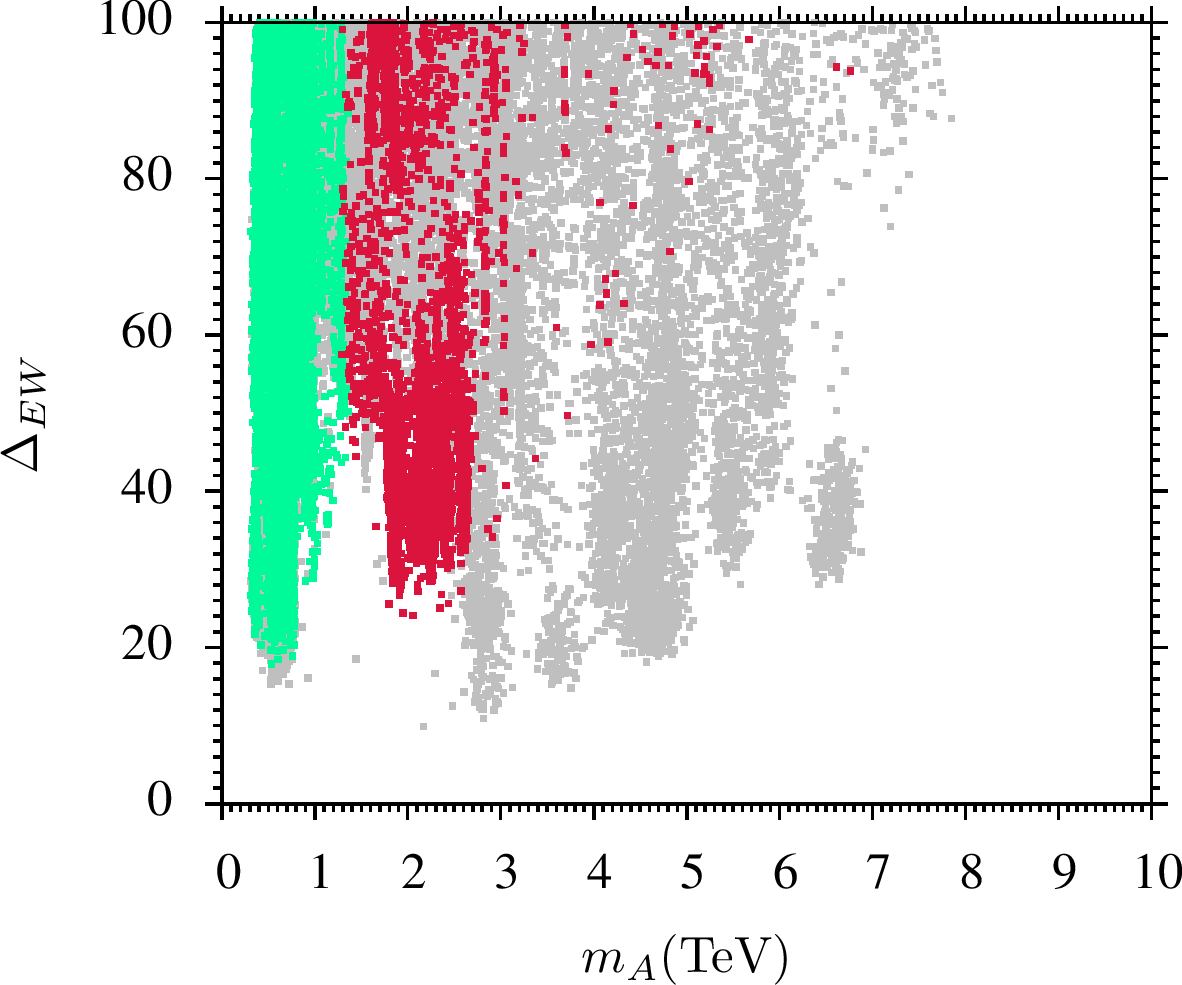}
}
\caption{$\Delta_{EW}$ vs. $\mu$, $m_{\tilde{t}_1}$, $A_t$ and $m_A$: Grey points satisfy the basic constraints \textbf{(\textit{\Rmnum{1}})}. Green points are a subset of grey points and satisfy the muon magnetic moment constraint \textbf{(\textit{\Rmnum{2}})}. Red points are a subset of green points and satisfy the Higgs coupling constraint \textbf{(\textit{\Rmnum{3}})} from CEPC (2 IP) only. 
}
\label{fig:muVSdeltaEW}
\end{figure}
In Figure~\ref{fig:muVSdeltaEW}, we show plots in $\mu-\Delta_{EW}$, $m_{\tilde t_1}-\Delta_{EW}$, $A_{t}-\Delta_{EW}$ and 
$m_A-\Delta_{EW}$ planes. In this figure, grey points satisfy the basic constraints (\textit{\Rmnum{1}}): REWSB; the lightest neutralino as an LSP condition; the sparticle mass bounds; the Higgs mass bound ($123\, {\rm GeV} \leq m_h \leq 127\, {\rm GeV}$); B-physics bounds; and fine-tuning bound ($\Delta_{\rm EW} \leq 100$). Green points are a subset of grey points and satisfy the muon anomalous magnetic moment constraint (\textit{\Rmnum{2}}). Red points are a subset of green points and satisfy the Higgs coupling constraint (\textit{\Rmnum{3}}). Here, to avoid the overlap of too many colors, when showing the effect of the Higgs coupling constraint (\textit{\Rmnum{3}}), we use the precisions of CEPC (2 IP) only.

In the top left panel, we see the obvious dependence of $\Delta_{EW}$ on 
$\mu$ as can be seen from Eq.~(\ref{eqn:Finetunning}). In this plot, the lowest value of $\Delta_{EW}$ we get is around 8 with $\mu \sim$ 100 GeV.
But when we apply constraint (\textit{\Rmnum{2}}), then $\Delta_{EW}$ goes up to 20 with $\mu \sim$ 140 GeV. For $\mu\sim$ 650 GeV, we have
$\Delta_{EW}$= 100. We also see that the constraints, which we have applied, do not have any preferred parameter space in this plane. All the
points are almost overlapped. Some spots with more grey points, less red points, and some void spots is just due to lack of statistics 
of data. By generating more points we may cover whole grey points by red points. Moreover, regions with large density
of points reflect our dedicated searches around some phenomenologically interesting points. This argument is also applied to all other figures in this paper. It is noted that since in
this plot $\mu$ is not that large, in order to have sizeable $\Delta a_{\mu}^{\rm SUSY}$ contributions, we should have either appropriately
large values of gaugino masses $M_{1,2}$ and $\tan\beta$ or small vales of SUSY particle as discussed in Section~\ref{subsec:g-2}. The is also the 
reason why $\mu$ goes up from 100 GeV to 140 GeV when we apply $\Delta a_{\mu}$ bound.

In the top right panel,
we see that for grey points $m_{\tilde t_1} \geq$ 0.5 TeV. This relatively large value of stop mass is due to the Higgs mass bounds.
We notice that for grey points, $\Delta_{EW}\sim$ 20 when $m_{{\tilde t}_1}\approx$ 1 TeV.  In this plot, we also see that grey, green and 
red points can be overlapped and constraints, which we have imposed, do not differentiate among these solutions very much.

Plot in the 
bottom left panel shows that we need $|A_{t}|\geq$ 5 TeV to be consistent with the basic constraints (\textit{\Rmnum{1}}). It is mainly because of the Higgs mass constraint. This plot shows that for 
$\Delta_{EW}\sim [20,100]$, we have $5 \,\rm TeV \leq |A_{t}| \leq  15 \, \rm TeV$ (red points).

In the bottom right panel, we see that
the Higgs coupling constraints, as is expected from the discussion in Section~\ref{subsec:HiggsCoupling}, are playing an important role.
Let us first comment on the relationship between $\Delta_{EW}$ and $m_{A}$ and for that we follow \cite{Bae:2014fsa}. From Eq.~(\ref{eqn:Chmu}), 
we have $C_{H_d} \equiv |m_{H_d}^2/({\rm tan}^2 \beta -1)|$. The tree-level value of $m_{A}$, in the case where $\mu^2 \sim -m^2_{H_{u}}$ and 
$m^2_{H_{d}}\gg -m^2_{H_{u}}$, can be given as 
\begin{equation}
m^2_{A}=m^2_{H_u}+m^2_{H_{d}}+2\mu^2 \approx m^2_{H_{d}}-m^2_{H_{u}} \sim m^2_{H_{d}}~.~\,
\end{equation}
From these equations one can have a rough estimation of $m_A$ in terms of $\Delta_{\rm EW}$ given as
\begin{equation}
m_{A} \leq m_{Z} \,\tan\beta \, \sqrt{\Delta_{\rm EW} (\rm max)}~,
\end{equation}
where $\Delta_{\rm EW} (\rm max)$ is the maximal fine-tuning one wants to allow. For example,
if we take $\Delta_{\rm EW}$= 20, with $\tan\beta$=10, we can have $m_{A}$ roughly as large as 4 TeV. This is why we see that 
in the plot  for a large range of $m_{A}$ values $[0.3,5]$ TeV, $\Delta_{EW}$ remains more or less
the same around 20. Some voids in the plots are just due to
lack of statistics. In this plot, a clear distinction between green and red points can be seen. Green points which are not covered by red points are in the $m_A$ mass range of $[0.3,1.2]$ TeV. On the other hand, the Higgs coupling ratios $k_{i}$ depend on $m_A$. Red points 
depict that Higgs coupling precision measurements at CEPC (2IP) can probe $m_A$ up to 1.2 TeV. We will show later that $m_A$ mass in the range of $[0.3,1.2]$ TeV 
can be probed by the precision measurements at the HL-LHC and ILC as well.

Here we want to comment on parameter space with low $m_A$. Current experiments are providing wealth of data and constraining parameter 
space of new physics such as SUSY, very effectively. For instance, the ATLAS direct searches for 
$g \, g\rightarrow A,H \rightarrow \tau^{+} \tau^{-}$ constrain $m_A$ and
$\tan\beta$ values \cite{Aad:2014vgg}. Since in our work we expect large deviations in the Higgs couplings such as $k_{b,\tau}$ for low $m_A$ values, we have to be careful about low $m_A$ solutions  because according to these bounds with small $m_A\sim$ 300 GeV, any point with 
 $\tan\beta\geq$ 18 is excluded. Since we are also considering scenario with low fine-tuning which implies low $\mu$ values.
As argured in~\cite{Bae:2015nva}, if $m_A$ is larger than $\mu$, then heavy Higgs bosons decay dominantly to charginos and neutralinos 
\cite{Bae:2014fsa}.
Thus, the  ATLAS bounds on $m_A$  can not be applied in this case. On the other hand, if $m_A < 2 \mu$, then one should apply ATLAS bounds 
on $m_A$ and $\tan\beta$. We will show later that for such a scenario, in our present scans, some part of parameter space is already excluded.

\subsection{Higgs Coupling Measurements}
\label{results_Higgs}

In this subsection, we study the Higgs coupling ratios $k_i$ in details. We apply only the basic constraints (\textit{\Rmnum{1}}) to study how Higgs coupling measurements can constrain the parameter space of EWSUSY in GmSUGRA.

\begin{figure}[htb!]
\centering
\subfigure{
\includegraphics[totalheight=5.5cm,width=7.cm]{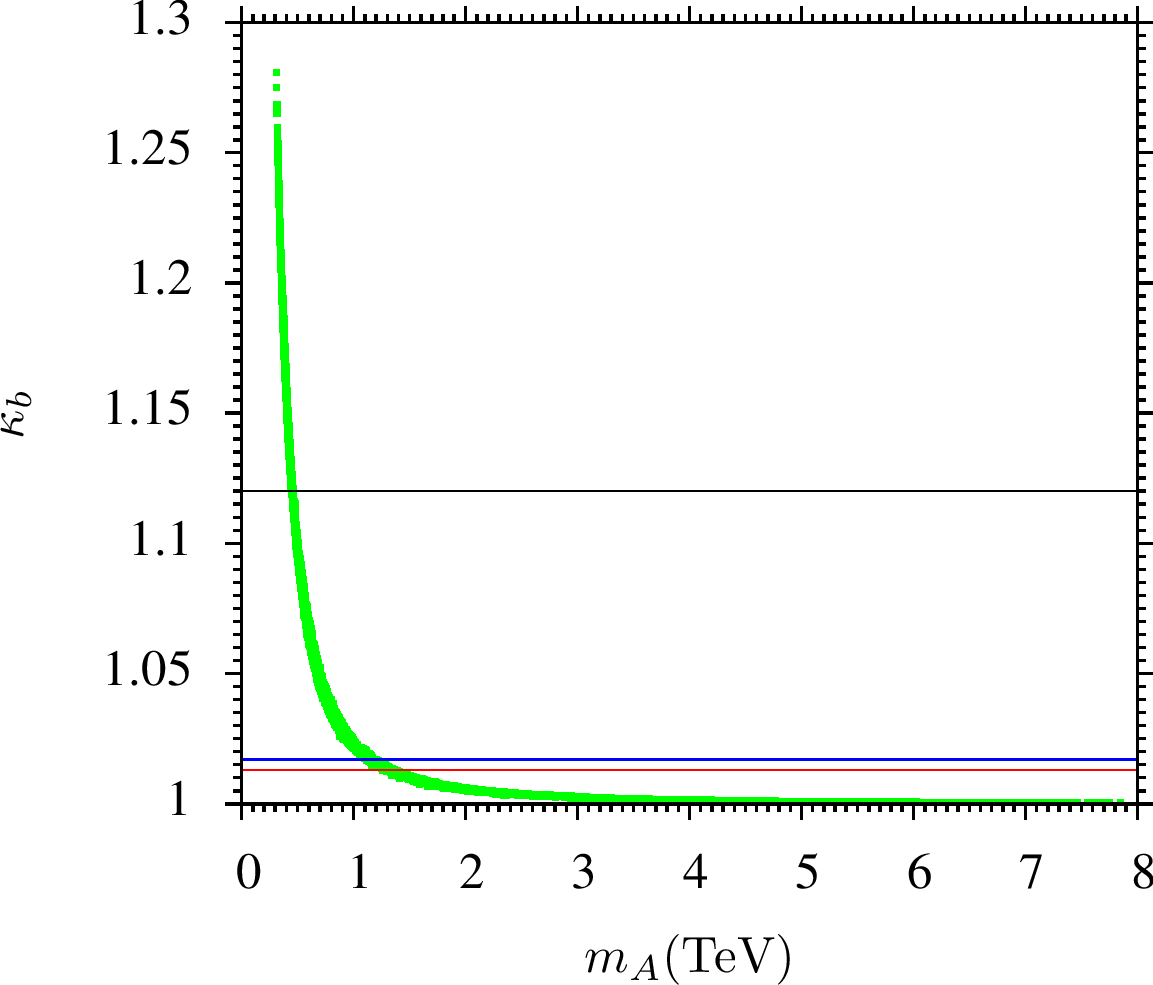}
}
\subfigure{
\includegraphics[totalheight=5.5cm,width=7.cm]{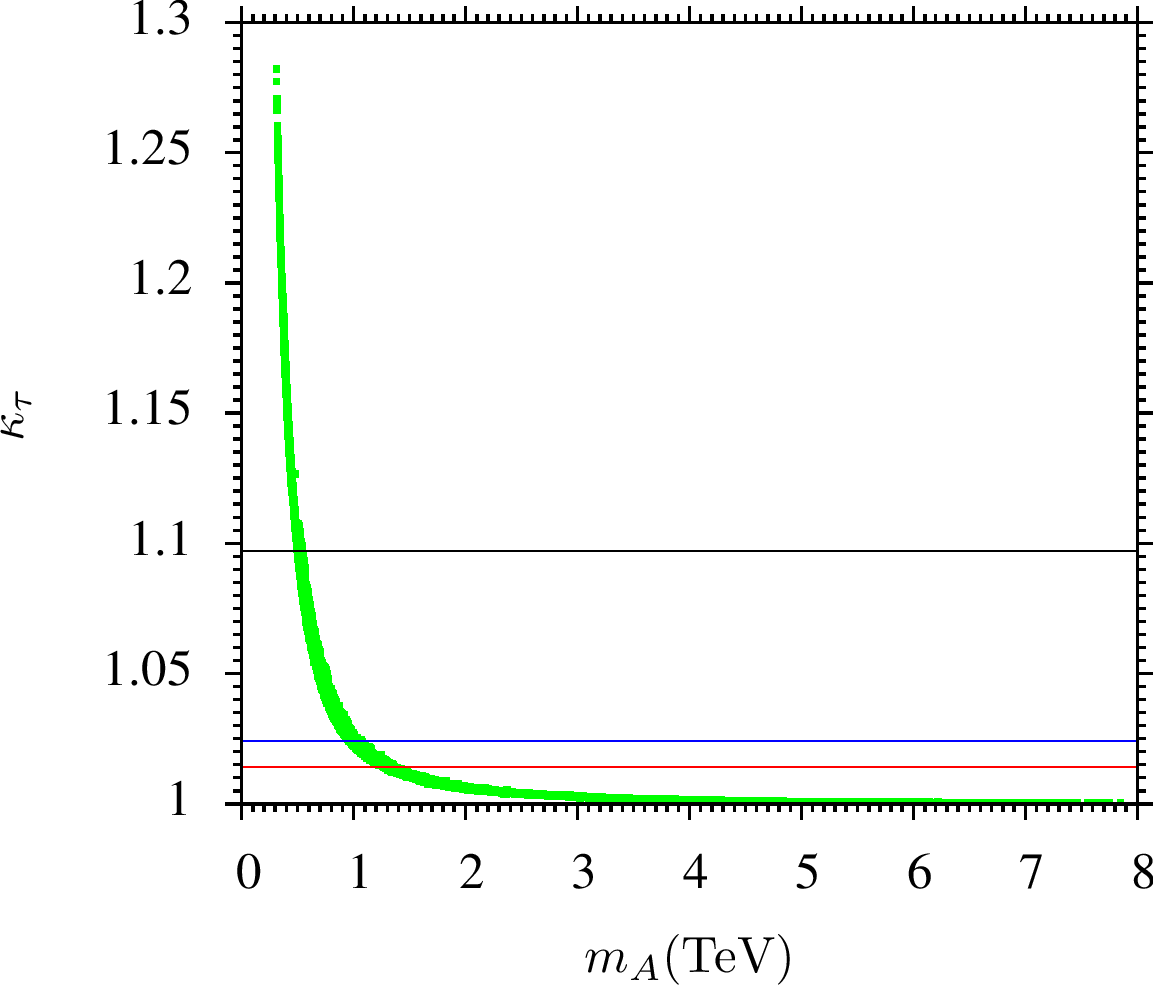}
}
\subfigure{
\includegraphics[totalheight=5.5cm,width=7.cm]{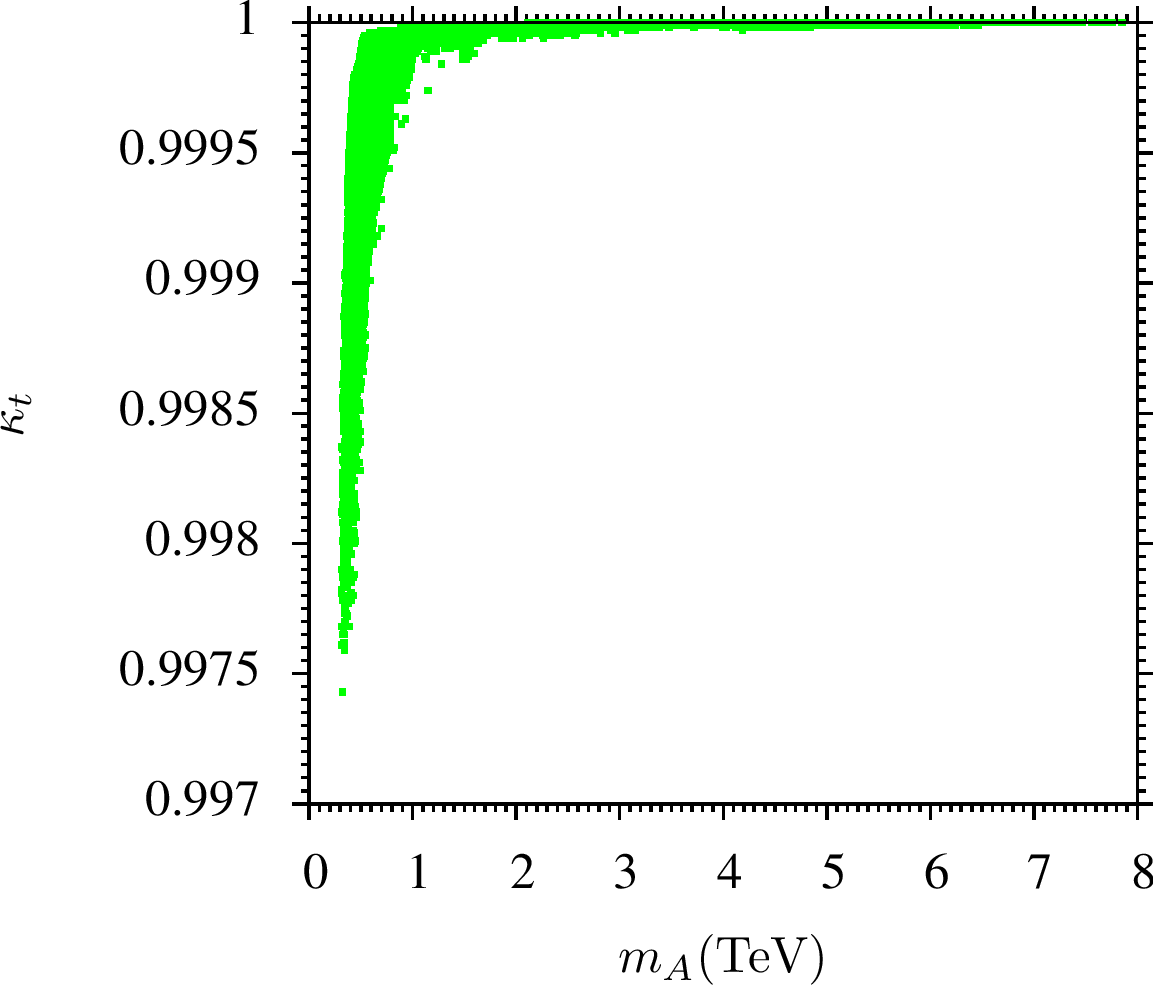}
}
\subfigure{
\includegraphics[totalheight=5.5cm,width=7.cm]{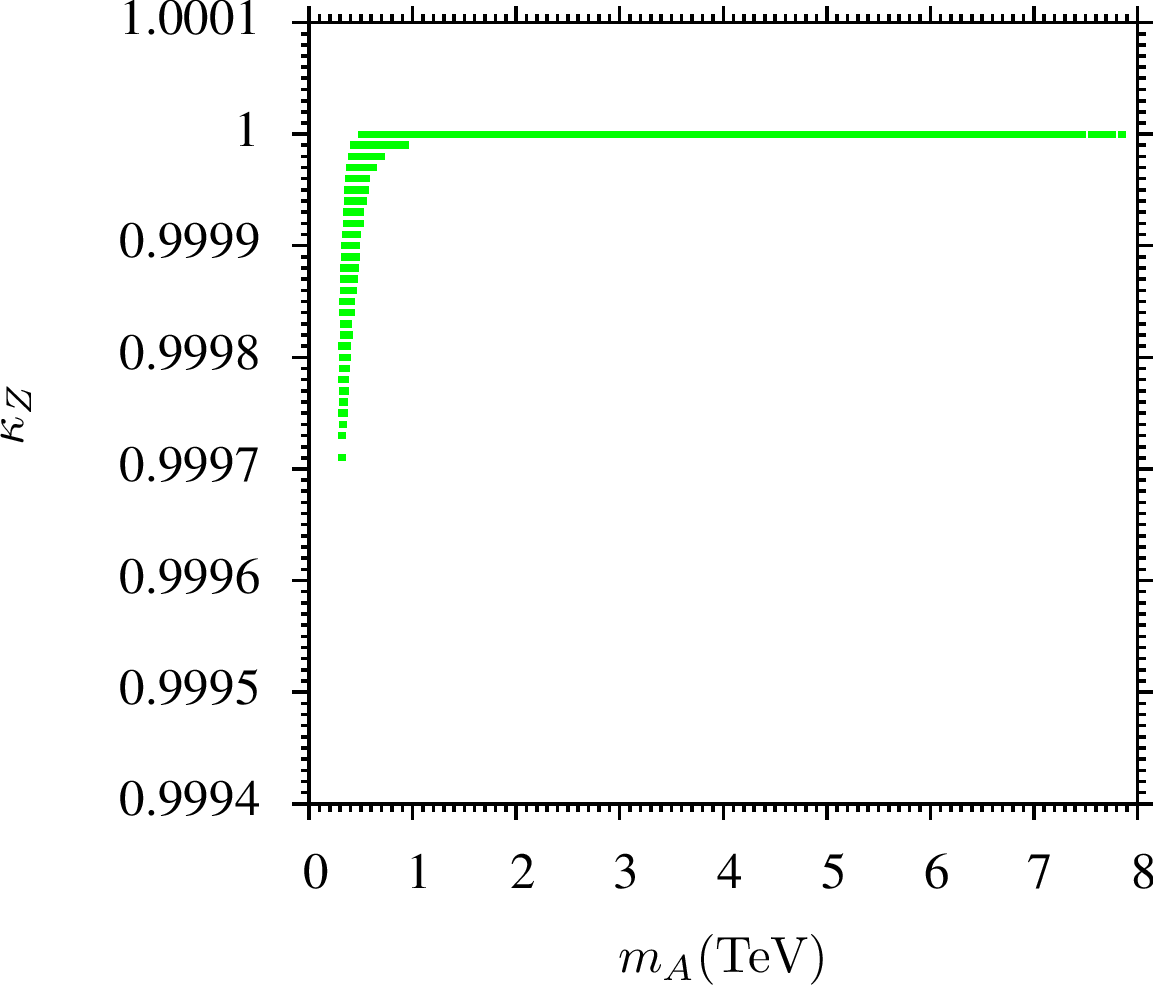}
}
\caption{$k_b$, $k_{\tau}$, $k_t$ and $k_Z$ vs. $m_A$: Green points satisfy the basic constraints \textbf{(\textit{\Rmnum{1}})}. The horizontal lines (if shown) in these plots label the precisions at different colliders: HL-LHC (black), ILC (blue) and CEPC (2 IP) (red) respectively.
}
\label{fig:kbVSmA}
\end{figure}

It is shown in Eqs.~(\ref{eqn:kv}), (\ref{eqn:kbktau}) and (\ref{eqn:kt}) that the Higgs coupling ratios $k_{W}$, $k_{Z}$, $k_b$, $k_{\tau}$ and $k_t$ are related to $m_A$. Among them, $k_{t}$, $k_{W}$ and $k_{Z}$ also have dependence on ${\rm tan}\beta$. In Figure \ref{fig:kbVSmA}, we show the parameter space in the $k_b$, $k_{\tau}$, $k_t$ and $k_Z$\footnote{We find the plot in the $k_W$ vs $m_A$ plane is very similar to the plot in the $k_Z$ vs $m_A$ plane, so we only show the $k_Z$ plot here.} versus $m_A$ planes. Green points in this figure satisfy the basic constraints (\textit{\Rmnum{1}}). The black, blue and red horizontal lines (if shown) in these plots correspond to the precisions at different colliders: HL-LHC, ILC and CEPC (2 IP), respectively.

These plots show the obvious dependence of Higgs coupling ratios $k_i$ on $m_A$. Firstly, about the sign of the deviations, $k_b$ and $k_{\tau}$ in these plots are always bigger than 1, while $k_t$, $k_W$ and $k_Z$ are smaller than 1. This is expected because according to
Eq.~(\ref{eqn:kbktau}), $k_b$ and $k_{\tau}$ have positive deviations, while from Eqs.~(\ref{eqn:kt}) and (\ref{eqn:kv}) the deviations of $k_t$, $k_W$ and $k_Z$ are all negative. 

Secondly, about the magnitude of the deviations, it is already shown that $k_{b,\tau}$, $k_t$ and $k_{W,Z}$ have the deviations around ${\cal O} ( 2 \frac{m_Z^2}{m_A^2} )$, ${\cal O} ( \frac{m_Z^2}{m_A^2} \frac{2}{\tan^2 \beta} )$, and ${\cal O} ( \frac{m_Z^4}{m_A^4} \frac{2}{\tan^2 \beta} )$, respectively. It is expected that with large $\tan\beta$, $k_{b,\tau}$ would have the maximal deviation. Due to the suppression of $\tan \beta$, the deviation in $k_t$ will be smaller compared with $k_{b, \tau}$. On the other hand, $k_{W,Z}$ has the smallest deviation
because of the suppressions of both $m_{A}^4$ and $\tan \beta$, which can be seen in the right bottom plot. The numerical results in our plots indicate that for a very small $m_A \approx 300$ GeV, the relative deviations in $k_{b,\tau}$, $k_t$ and $k_{W,Z}$ can be 28\%, 0.3\% and 0.03\% respectively. 

Thirdly, $m_A$ can be constrained by the Higgs coupling measurements. When $m_A \leq 1$ TeV, $k_b$ and $k_{\tau}$ are very sensitive to $m_A$. The precision of $k_b$ ( $k_{\tau}$) at the HL-LHC can constrain $m_A$
to be above 0.4 (0.5) TeV. The precision at the ILC can constrain $m_A$ to be above 1.1 (0.9) TeV. The precision at the CEPC (2 IP) can constrain the $m_A$ to above 1.2 (1.1) TeV. However, the deviations in $k_t$, and $k_{W,Z}$ are so small that even the precision at CEPC (2 IP) cannot constrain $m_A$. To be able to constrain $m_A$, the precision of $k_t$ needs to be better than 0.3\% and the precision of $k_{W,Z}$ need to be better than 0.03\%. It should be noted that a future $e^+e^-$ collider with more IP and longer running time, can offer a higher integrated luminosity and thus the better precisions. For example, the FCC-ee (4 IP) can have a larger integrated luminosity of 10 ${\rm ab}^{-1}$, where the precisions of $k_b$ and $k_{\tau}$ can be 0.88\% and 0.94\% respectively (see Table \ref{tab:HiggsCplP}). These precision 
measurements of $k_b$ and $k_{\tau}$ can constrain $m_A$ to above 1.55 TeV and 1.5 TeV respectively. But the precision measurements of $k_t$ and $k_{W,Z}$ at FCC-ee (4 IP) are still not good enough to constrain $m_A$.

\begin{figure}[htb!]
\centering
\subfigure{
\includegraphics[totalheight=5.5cm,width=7.cm]{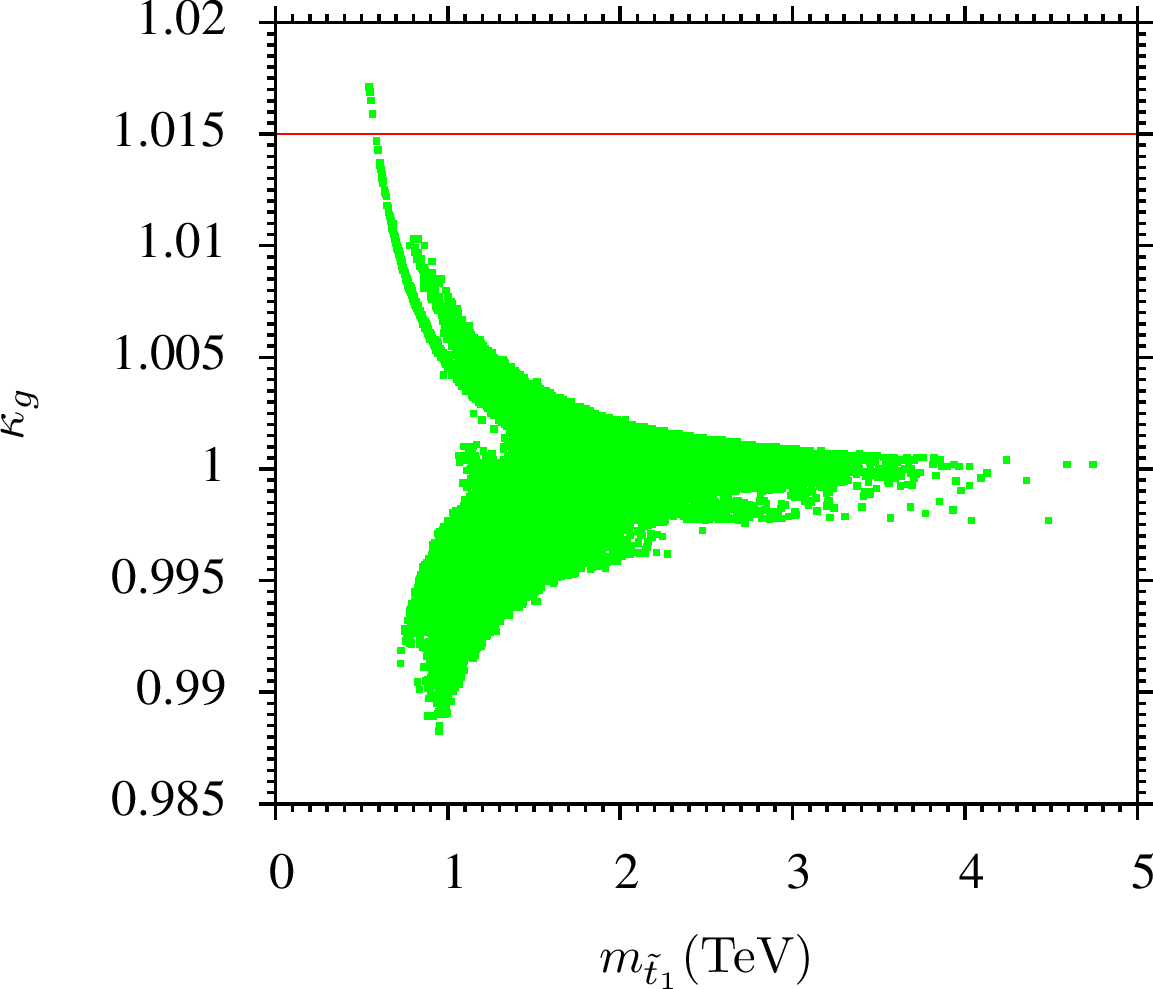}
}
\subfigure{
\includegraphics[totalheight=5.5cm,width=7.cm]{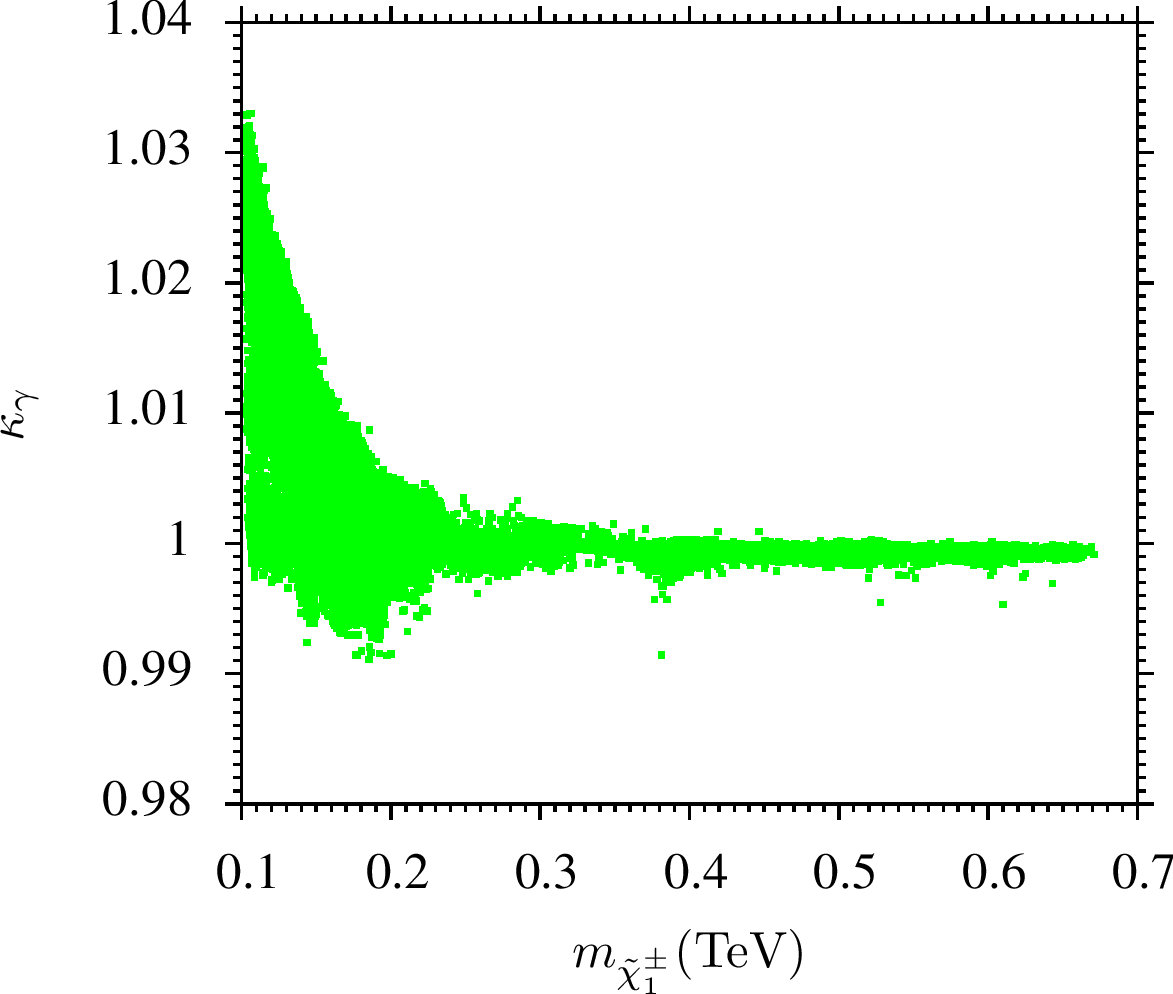}
}
\caption{$k_g$ vs. $m_{\tilde{t}_1}$ and $k_{\gamma}$ vs. $m_{\tilde{\chi}_1^{\pm}}$: Color coding and the horizontal line are the same as Figure \ref{fig:kbVSmA}.
}
\label{fig:kgVSmStop}
\end{figure}

In Figure \ref{fig:kgVSmStop}, we show plots in $k_g-m_{\tilde{t}_1}$ and $k_\gamma-m_{\tilde{\chi}_1^{\pm}}$ planes. The color coding and the horizontal lines are the same as Figure \ref{fig:kbVSmA}.

In SUSY, the dominant contribution to $hgg$ coupling can come from stops.
The left plot presents the quantitative dependence of Higgs coupling ratio $k_g$ on $m_{\tilde{t}_1}$. It can be seen that for $m_{\tilde t_1} \sim$ 2.5 TeV to 5 TeV, the deviation from the SM $hgg$ coupling is almost negligible. But as the $m_{\tilde t_1}$ decreases, the deviation starts growing. $k_g$ is 
getting greater than 1 and also getting less than one. Around $m_{\tilde t_1} \sim$ 0.5 TeV, $k_g$ is around 1.017 (1.7\% deviation from the SM
coupling) while for
$m_{\tilde t_1} \sim$ 0.9 TeV, $k_g \sim$ 0.988. This can be understood from Eq.~(\ref{eqn:kg}). We note that for points with $k_g \geq$ 1.01, not only stops are
light but also scalar trilinear coupling $A_{t}$ is positive. Since we have considered ${\rm sgn}(\mu) >$ 0, we get $X_{t}=A_{t}-\mu/\tan\beta \leq A_{t}$. On the other hand, for the points
with $k_g <$ 1, we notice that $A_{t}$ is negative. This means $|X_{t}|\geq |A_{t}|$. Even though in Eq.~(\ref{eqn:kg}), we have 
$X_{t}^2$ term, but the magnitude of the contribution is different, so with small stops and small negative contribution from $X_{t}^2$,
one can have $k_g > 1$. On the other hand with relatively large stop masses, and relatively large negative contribution from
$X_{t}^2$ term, one can get $k_g <$ 1. From the horizontal line, we see that the precision of $k_g$ at CEPC (2 IP) can bound $m_{\tilde{t}_1}$ to around 600 GeV.

Like $k_g$, $k_{\gamma}$ is also a loop induced coupling. Higgs boson can couple to $\gamma \gamma$ pair via loops of all the SM charged
particles which are quarks, lepton and $W^{\pm}$ and through squarks, slepton, $H^{\pm}$ and $\tilde \chi_{1}^{\pm}$.
Since we are considering parameter space with low fine-tuning, as is shown in Figure~\ref{fig:muVSdeltaEW} that $\mu$ should be relatively small. We expect to have light higgsinos.
In the right plot, we see that for small values of chargino, the values of $k_{\gamma}$ can be large. As the chargino
mass increases, deviation goes down. In our present scans, the deviation in $k_{\gamma}$ can be up to 3\% with 
$m_{\tilde{\chi}_1^{\pm}} \approx 100$ GeV. We also see that there are a few points with $k_{\gamma}$ below 1. This is because for these points the contribution of
$\tilde b_1$ and $\tilde \tau$ is negative.

It is worth noting that a better precision 1.1\% of $k_g$ at FCC-ee (4 IP) can constrain $m_{\tilde{t}_1}$ to be above 700 GeV. Besides, the precision of $k_\gamma$ at FCC-ee (4 IP) can be 1.7\%, which can constrain $m_{\tilde{\chi}_1^{\pm}}$ to above 150 GeV.

\subsection{The muon magnetic moment measurement}
\label {results_Amu}

In this subsection, we use the measurement of the muon magnetic moment to constrain the related sparticle masses.

\begin{figure}[htb!]
\centering
\subfigure{
\includegraphics[totalheight=5.5cm,width=7.cm]{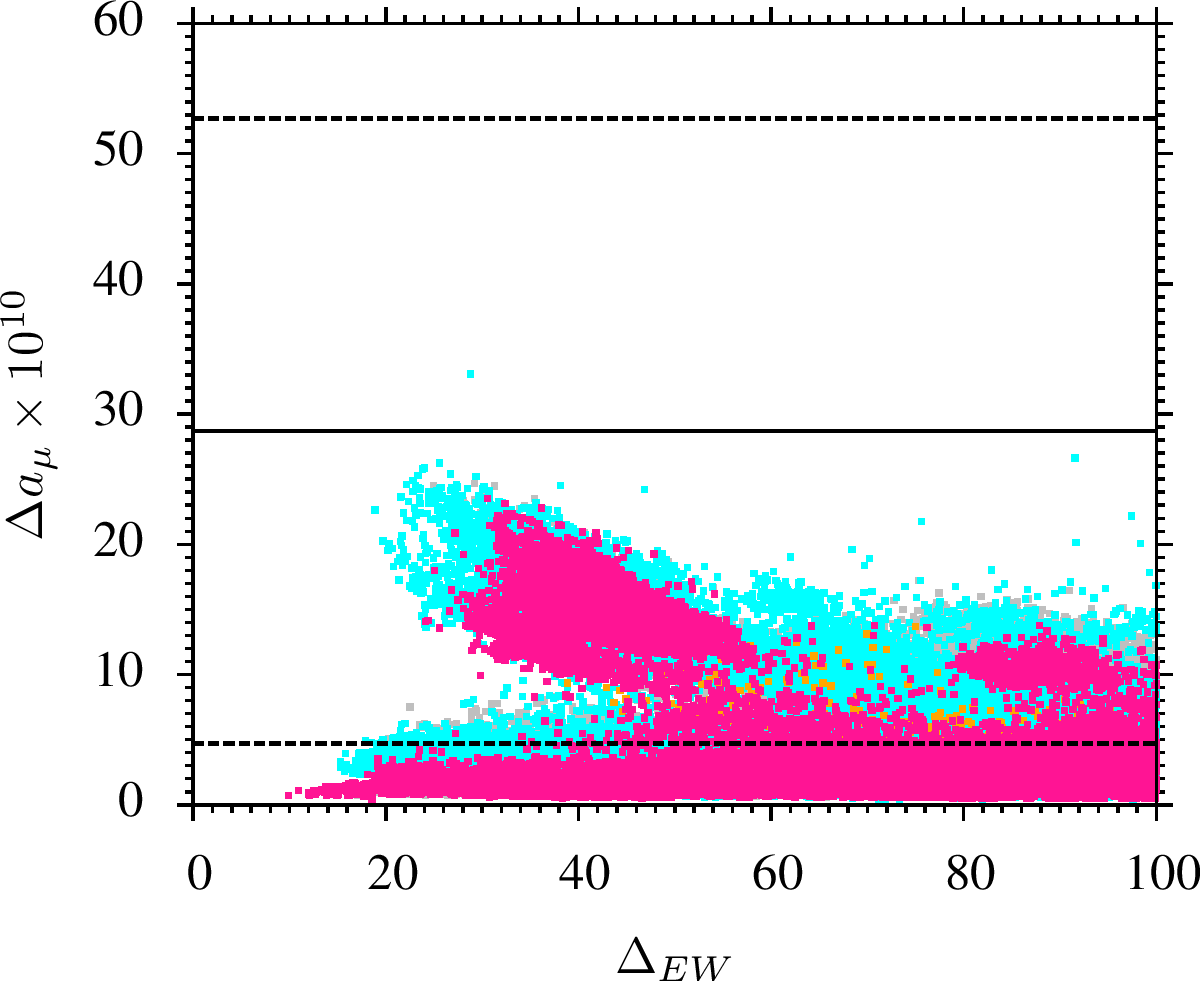}
}
\caption{$\Delta a_{\mu}$ vs. $\Delta_{EW}$: Grey points satisfy the basic constraints \textbf{(\textit{\Rmnum{1}})}. Cyan, orange and red points are subsets of grey points and satisfy the Higgs coupling constraint \textbf{(\textit{\Rmnum{3}})} from HL-LHC, ILC, and CEPC (2 IP), respectively. The solid horizontal line displays the central value of $\Delta a_{\mu}$ from the muon magnetic moment experiment, while the dash lines show the 3-$\sigma$ values.
}
\label{fig:AmuVSdeltaEW}
\end{figure}

In Figure \ref{fig:AmuVSdeltaEW}, we display a plot in the $\Delta a_{\mu} - \Delta_{\rm EW}$ plane. 
In this plot, grey points satisfy the basic constraints (\textit{\Rmnum{1}}). Cyan, orange and red points are subsets of grey points and satisfy the Higgs coupling constraint (\textit{\Rmnum{3}}) from HL-LHC, ILC, and CEPC (2 IP), respectively. The solid horizontal line labels the central value of $\Delta a_{\mu}$ from Eq.~(\ref{eqn:ThDeltaAmu}), while the dash lines show the 3-$\sigma$ values.
We see that because of the basic constraints (\textit{\Rmnum{1}}), almost all of the points are below the solid line except one single
point. This shows that if we just keep generating more data, we can have more solutions there. Similar argument can also be applied on the 
isolated points below the solid line. One can see a rising trend in points with $\Delta_{EW}$ from 40 to 20. It is just an artefact of our dedicated
searches for low $\Delta_{EW}$ values. In fact, it can be seen from the figure that more points can be generated for $\Delta_{EW}\sim [20,100]$ 
with appropriate contributions to $\Delta a_{\mu}$. We also notice that the effects of the application of the above constraints are nearly 
indistinguishable and the points satisfying various constraints are nearly overlapped. But our present scan shows that the parameter space can still be constrained by the combination of muon magnetic moment and Higgs coupling measurements effectively. Using the precision at CEPC (2 IP) one may bound the $\Delta a_{\mu}$ to be between $(5 \sim 25)\times 10^{-10}$ while $\Delta_{EW} \geq 30$.

\begin{figure}[htb!]
\centering
\subfigure{
\includegraphics[totalheight=5.5cm,width=8.0cm]{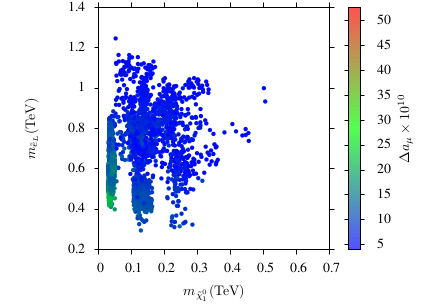}
}
\subfigure{
\includegraphics[totalheight=5.5cm,width=8.0cm]{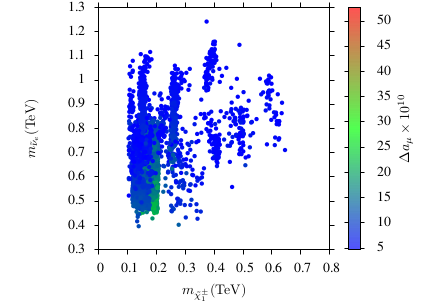}
}
\subfigure{
\includegraphics[totalheight=5.5cm,width=8.0cm]{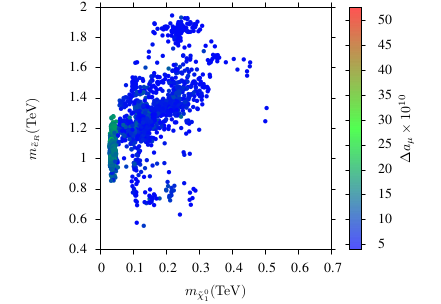}
}
\caption{$m_{\tilde{e}_{R,L}}$ vs. $m_{\tilde{\chi}_1^0}$ and $m_{\tilde{\nu}_{e}}$ vs. $m_{\tilde{\chi}_1^{\pm}}$: Points satisfy both the basic constraints \textbf{(\textit{\Rmnum{1}})}, the muon magnetic moment constraint \textbf{(\textit{\Rmnum{2}})} and the Higgs coupling constraint (\textit{\Rmnum{3}}) from the CEPC (2 IP) only. Color represents the $\Delta a_{\mu}$ values.
}
\label{fig:AmuVSer}
\end{figure}

In SUSY, as we have discussed in Section \ref{subsec:g-2}, the contributions to the muon anomalous magnetic moment $\Delta a_{\mu}$ mainly come from the 1-loop diagrams of both $\tilde{e}_{R,L}-{\tilde{\chi}_1^0}$ and $\tilde{\nu}-{\tilde{\chi}_1^{\pm}}$. In Figure \ref{fig:AmuVSer}, we show plots in $m_{\tilde{e}_{L, R}} - m_{\tilde{\chi}_1^0}$ and $m_{\tilde{\nu}} - m_{\tilde{\chi}_1^{\pm}}$ planes. 
In these plots, all points satisfy the basic constraints (\textit{\Rmnum{1}}), the muon magnetic moment constraint (\textit{\Rmnum{2}}) and the Higgs coupling constraint (\textit{\Rmnum{3}}). To avoid the overlap of too many colors, when showing the effect of the Higgs coupling constraint (\textit{\Rmnum{3}}), we use the precisions of CEPC (2 IP) only. The vertical color bars represent the spread of $\Delta a_{\mu}$ values in these plots.
 
In the top left plot,  
we see that by applying the above constraints, the left handed slepton mass can be constrained to $[0.3, 1.2]$ TeV. But to have 
sizeable contribution to $\Delta a_{\mu}$ (greenish blue points), the left-handed slepton mass needs to be in the range of $[0.35,0.9]$ TeV and neutralino mass in the range of $\leq$ 0.17 TeV. The bright green points corresponding to large values of $\Delta a_{\mu} \geq  20 \times 10^{-10}$ have narrow neutralino mass range of $[0.03,0.1]$ TeV and $m_{\tilde e_{L}} \sim [0.4,0.5]$ TeV. The top right plot indicates that for the blue, greenish blue and bright green points, sneutrino have more or less same mass ranges as the the left-handed slepton, while chargino mass are in the range of $[0.1,0.7]$ TeV, $[0.15,0.25]$ TeV and $[0.16,0.22]$ receptively.

In the bottom panel, we see that the lightest neutralino mass can be restricted 
below 0.5 TeV and the right-handed selectron mass is in the range of $[0.5,2]$ TeV. It is also clear from this plot that most of the points have
$\Delta a_{\mu}$ from $5\times 10^{-10}$ to $10 \times 10^{-10}$ (blue points). But one can also see greenish blue points in the region where neutralino mass is in the range of $[0.04,0.3]$ TeV and right-handed slepton mass in the range of $[0.7,1.6]$ TeV. Bright green points exist in the region with $m_{\tilde e_{R}}$ mass in the range of $[0.8,1.3]$ TeV and neutralino mass 
$\leq$ 0.15 TeV. Comparing with Figure~\ref{fig:AmuVSdeltaEW}, it can be understood that these points correspond to $\Delta_{EW}\sim$ 30-40.

\subsection{Collider Phenomenology}
\label {results_Others}

In this section, we show some model parameters and interesting sparticle masses after applying all constraints and discuss the possible collider phenomenology.

\begin{figure}[htb!]
\centering
\subfigure{
\includegraphics[totalheight=5.5cm,width=7.cm]{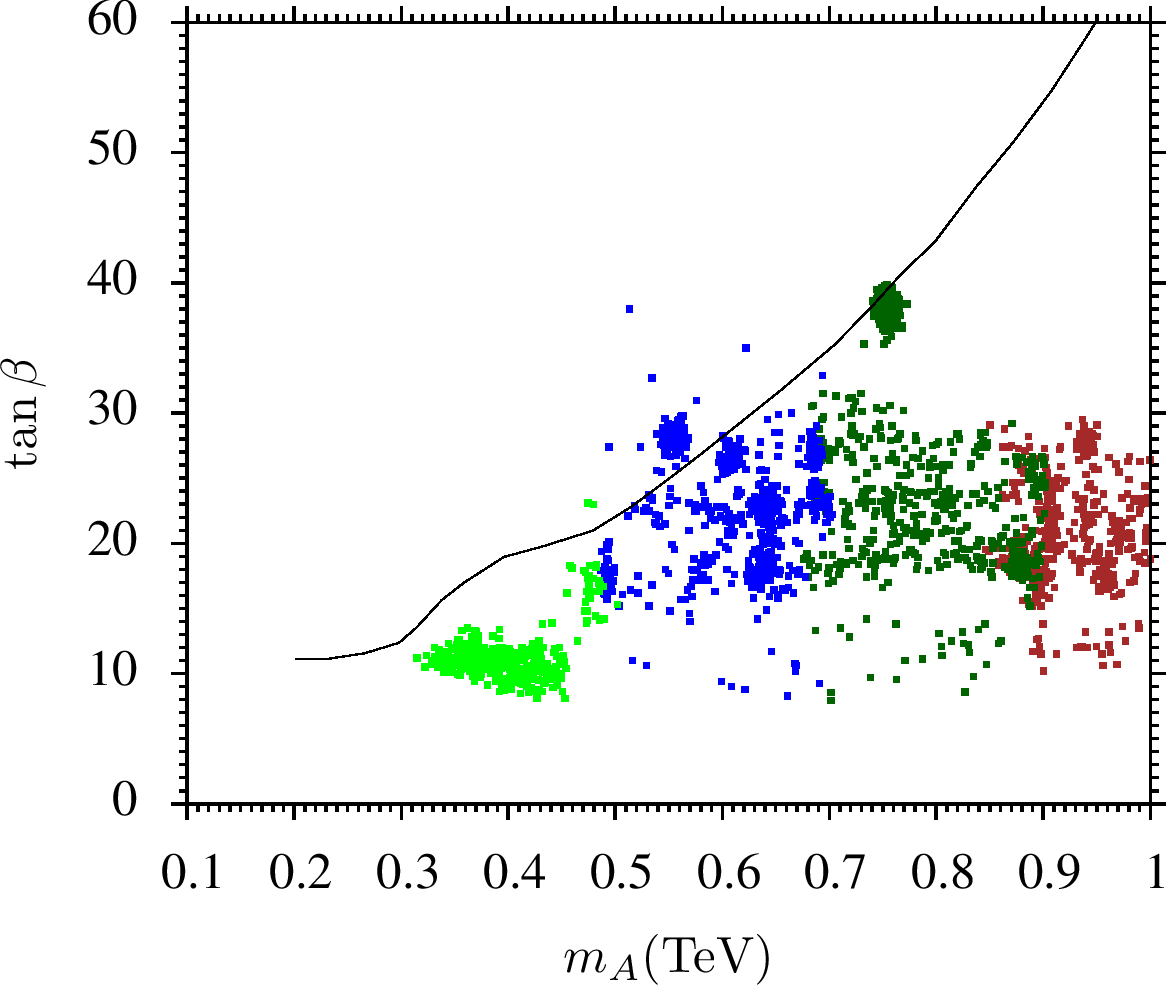}
\includegraphics[totalheight=5.5cm,width=7.cm]{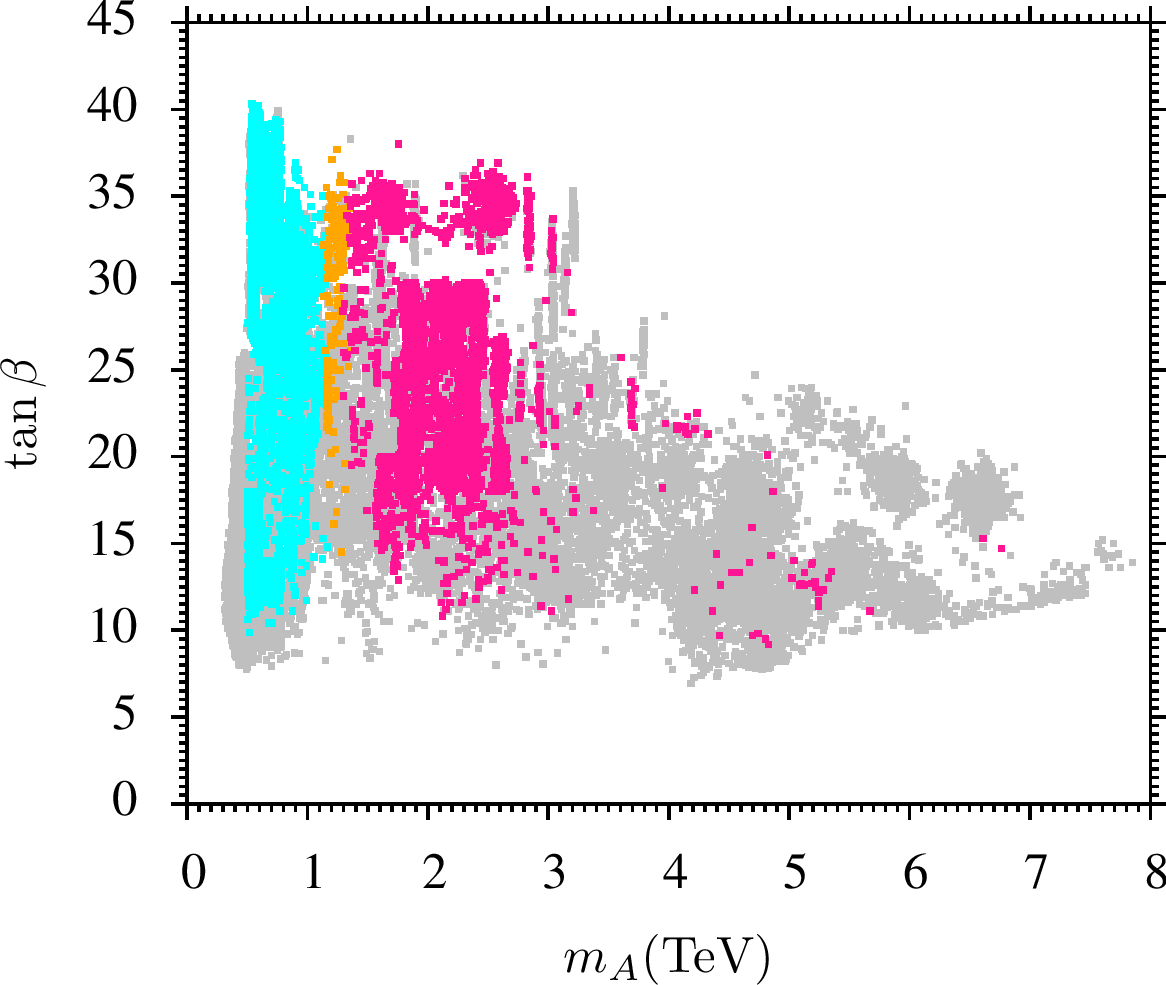}
}
\caption{${\rm tan}\beta$ vs. $m_A$: In the left plot, all points satisfy the basic constraints \textbf{(\textit{\Rmnum{1}})} and $m_A\leq 2 \mu$; green, blue, dark green and brown points represent $k_{b} \geq 1.1$,
$1.05 \le k_{b}\le 1.1$, $1.03 \le k_{b}\le 1.05$ and $k_{b} \le1.03$ respectively; black curve shows the $1-\sigma$ bound from the ATLAS direct heavy Higgs search~\cite{Aad:2014vgg}.
In the right plot, grey points satisfy the basic constraints \textbf{(\textit{\Rmnum{1}})}; cyan, orange and red points are subsets of grey points, and satisfy both the muon magnetic moment constraint \textbf{(\textit{\Rmnum{2}})} and the Higgs coupling constraint \textbf{(\textit{\Rmnum{3}})} from HL-LHC, ILC, and CEPC (2 IP), respectively.
}
\label{fig:tanBetaVSmA}
\end{figure}

In Figure \ref{fig:tanBetaVSmA}, we show plots in the ${\rm tan}\beta - m_A$ plane. In the left plot, all points satisfy the basic constraints (\textit{\Rmnum{1}}) and $m_A\leq 2 \mu$. Green, blue, dark green and brown points represent $k_{b} \geq 1.1$,
$1.05 \le k_{b}\le 1.1$, $1.03 \le k_{b}\le 1.05$ and $k_{b} \le1.03$, respectively. Black curve shows the 1 $\sigma$ 
bounds in the ${\rm tan}\beta - m_A$ plane from the ATLAS direct heavy Higgs search~\cite{Aad:2014vgg}.

As we have discussed earlier in the Section~\ref{results_Naturalness}, these are the points where $m_A$ can decay into the SM modes
such as $\tau^{+} \tau^{-}$. Here we restrict $m_A$ up to 1 TeV to compare our results with Figure 10a of~\cite{Aad:2014vgg}. Since the ATLAS search relies on the $\tau^{+} \tau^{-}$ final states, it is understood that the ATLAS direct search mainly exclude the top left region in this plot where $\tan\beta$ is large which enhances the branching ratio of $A \rightarrow \tau^{+} \tau^{-}$, and $m_A$ is small which can give large production of heavy Higgs. However, in this study we bound $m_A$ by the Higgs coupling measurement which do not have too much $\tan\beta$ restriction. Therefore, our bound can exclude the region in this plot with small $\tan\beta$ as well. In this sense, the Higgs coupling measurement is complementary to the direct search of heavy Higgs when constraining SUSY.

It can be seen clearly 
that in our present scans, only very small part of our data is excluded by the ATLAS results. Interestingly, most points with large 
deviations in $k_b$, shown as green points in the plot, survive because of the corresponding small $\tan\beta$ and only a few points of this kind can be excluded by the direct searches.
We hope in near future the remaining part of this parameter space
will soon be probed and we can have better understanding about the Higgs couplings. Moreover, we also notice that dark green points which satisfy
CEPC (2IP) bounds, can also be within the range of ATLAS heavy Higgs searches. 

In the right plot, we show our full data after the application of all constraints.
The grey points satisfy the basic constraints (\textit{\Rmnum{1}}); the cyan, orange and red points are subsets of grey points, and satisfy both the muon magnetic moment constraint (\textit{\Rmnum{2}}) and the Higgs coupling constraint (\textit{\Rmnum{3}}) from HL-LHC, ILC, and CEPC (2 IP), respectively. The combination of the muon magnetic moment and the Higgs coupling precision measurements can constrain ${\rm tan}\beta$ and $m_A$ effectively. We see that due to the Higgs coupling precisions at HL-LHC, ILC and CEPC (2 IP), the $m_A$ can be bounded to be above around 0.5 TeV, 1.1 TeV and 1.2 TeV respectively. The $\tan \beta$ is bounded in the range between 10 and 35. We comment here that since the ${\rm BR}(B_s \rightarrow \mu^+ \mu^-) \propto ( \frac{\tan^6 \beta}{m_A^4} )$, our B-physics bounds in the basic constraints (\textit{\Rmnum{1}}) also exclude a part of parameter space in this plane. 

\begin{figure}[htb!]
\centering
\subfigure{
\includegraphics[totalheight=5.5cm,width=7.cm]{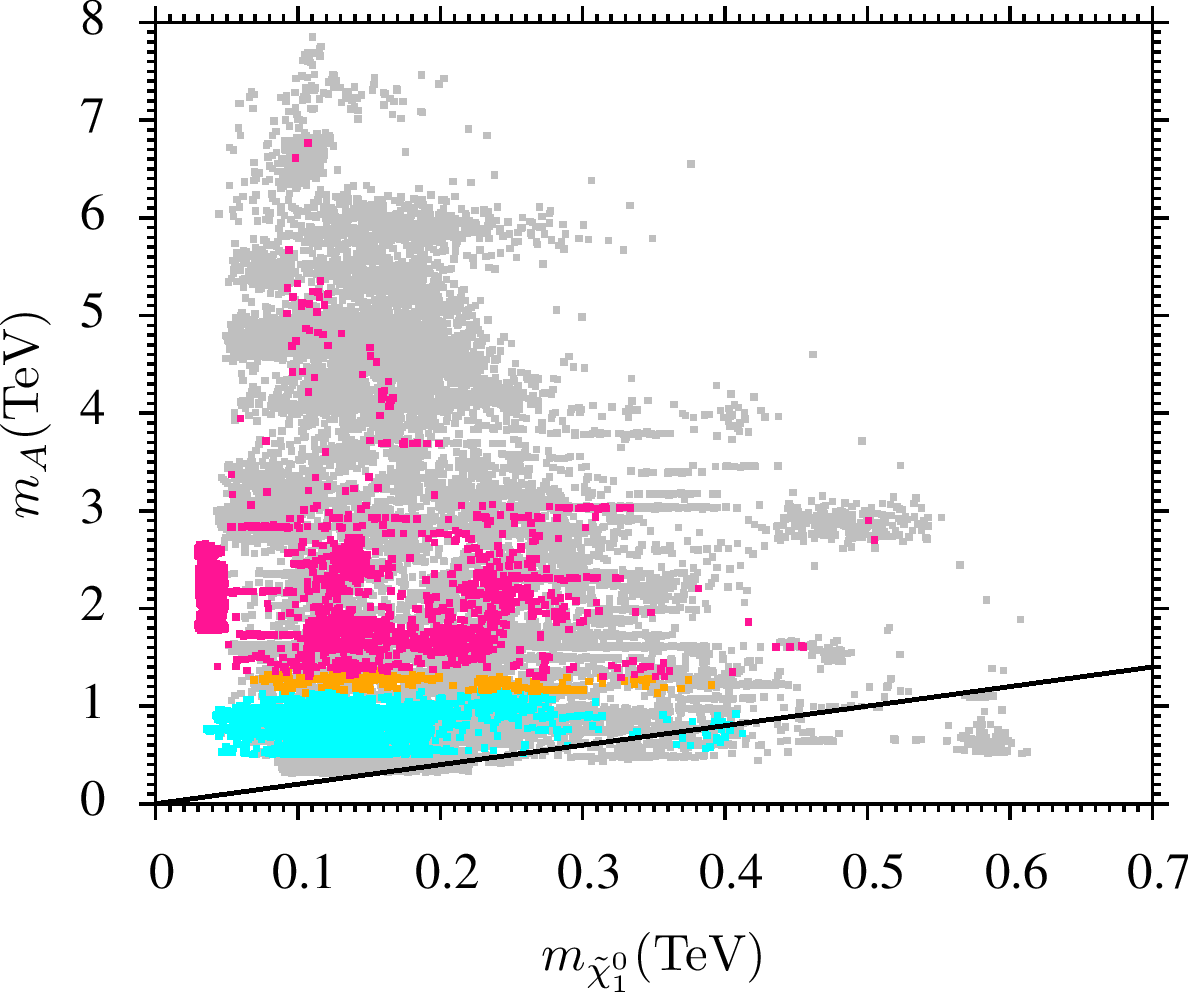}
}
\subfigure{
\includegraphics[totalheight=5.5cm,width=7.cm]{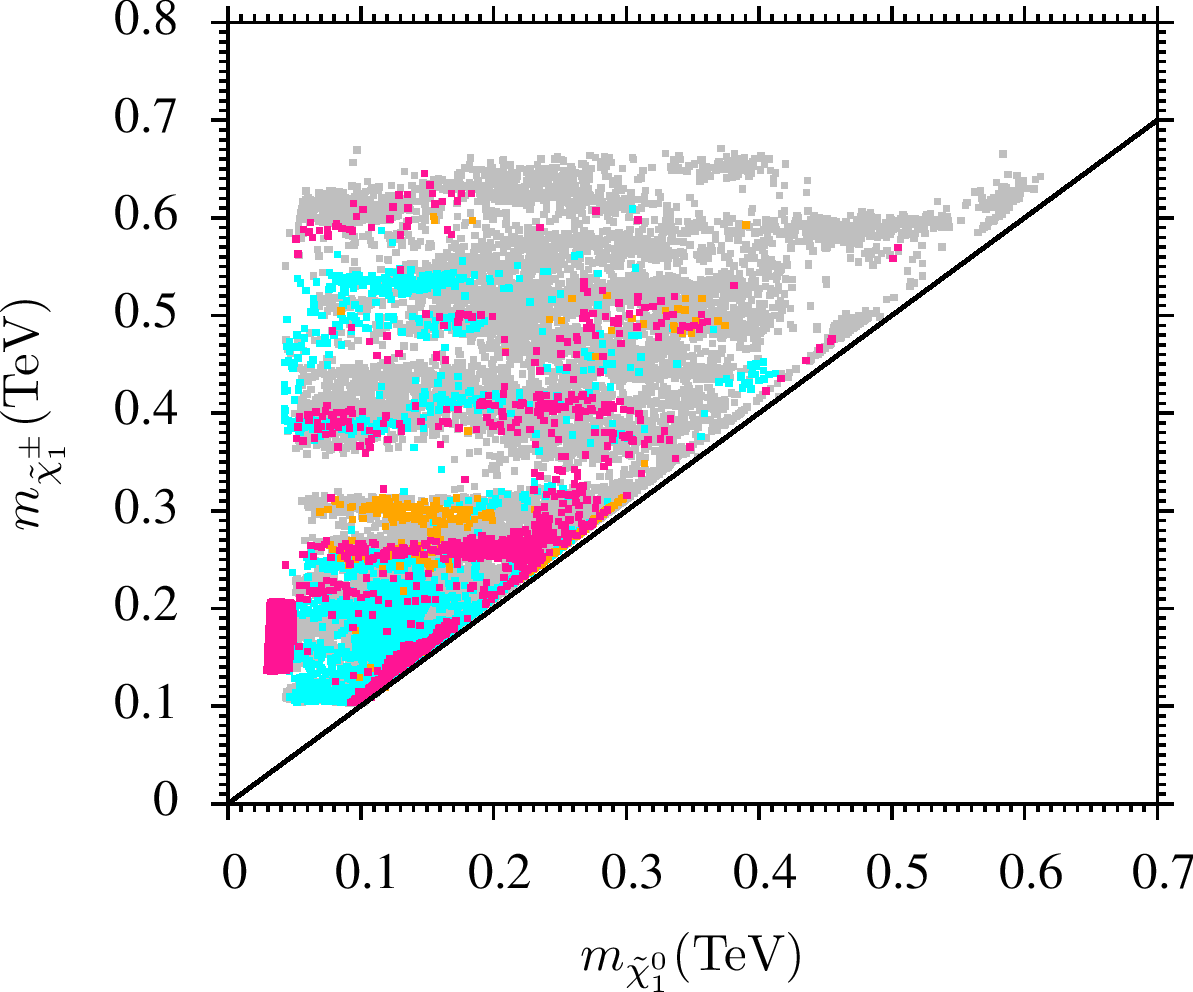}
}
\subfigure{
\includegraphics[totalheight=5.5cm,width=7.cm]{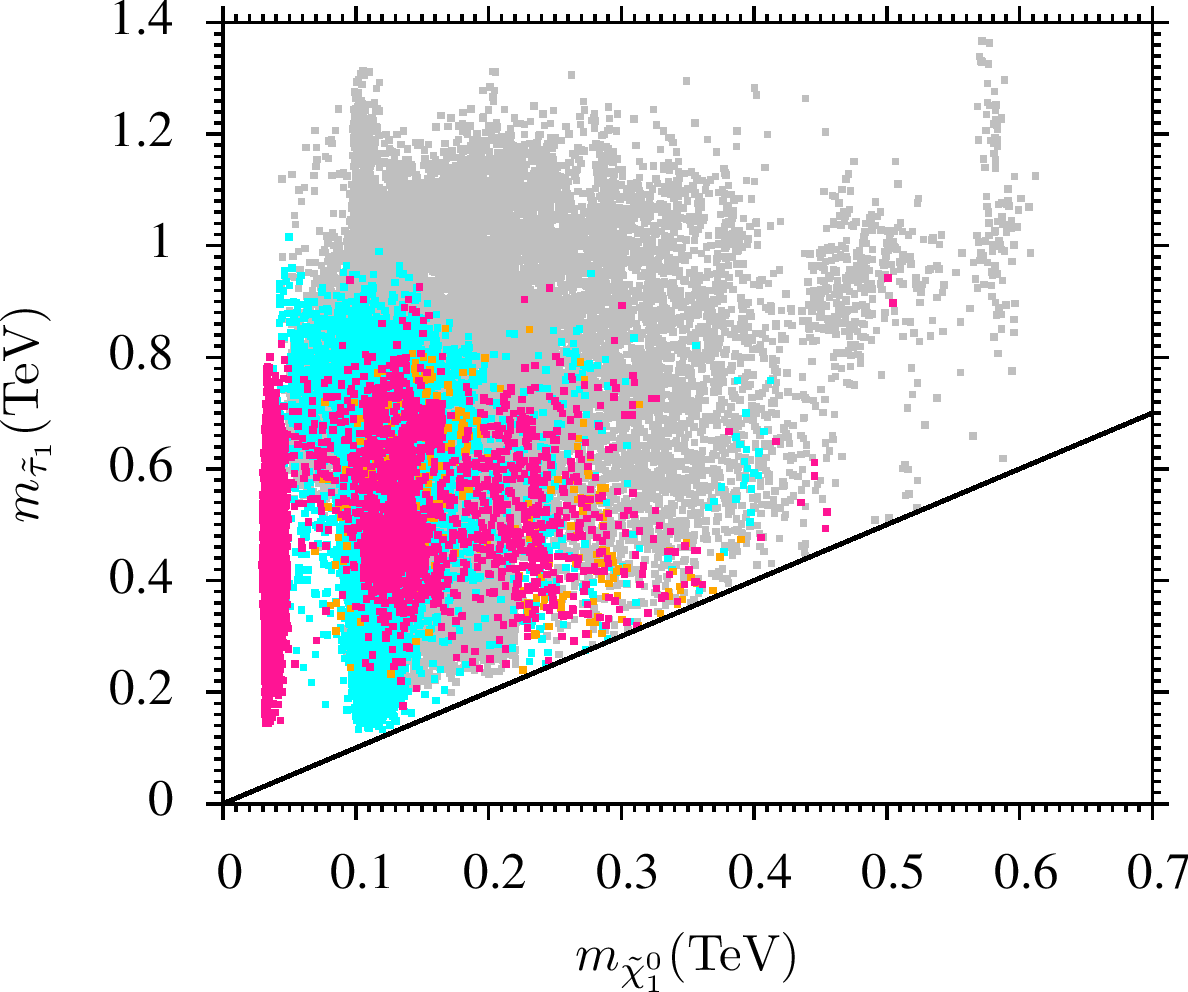}
}
\subfigure{
\includegraphics[totalheight=5.5cm,width=7.cm]{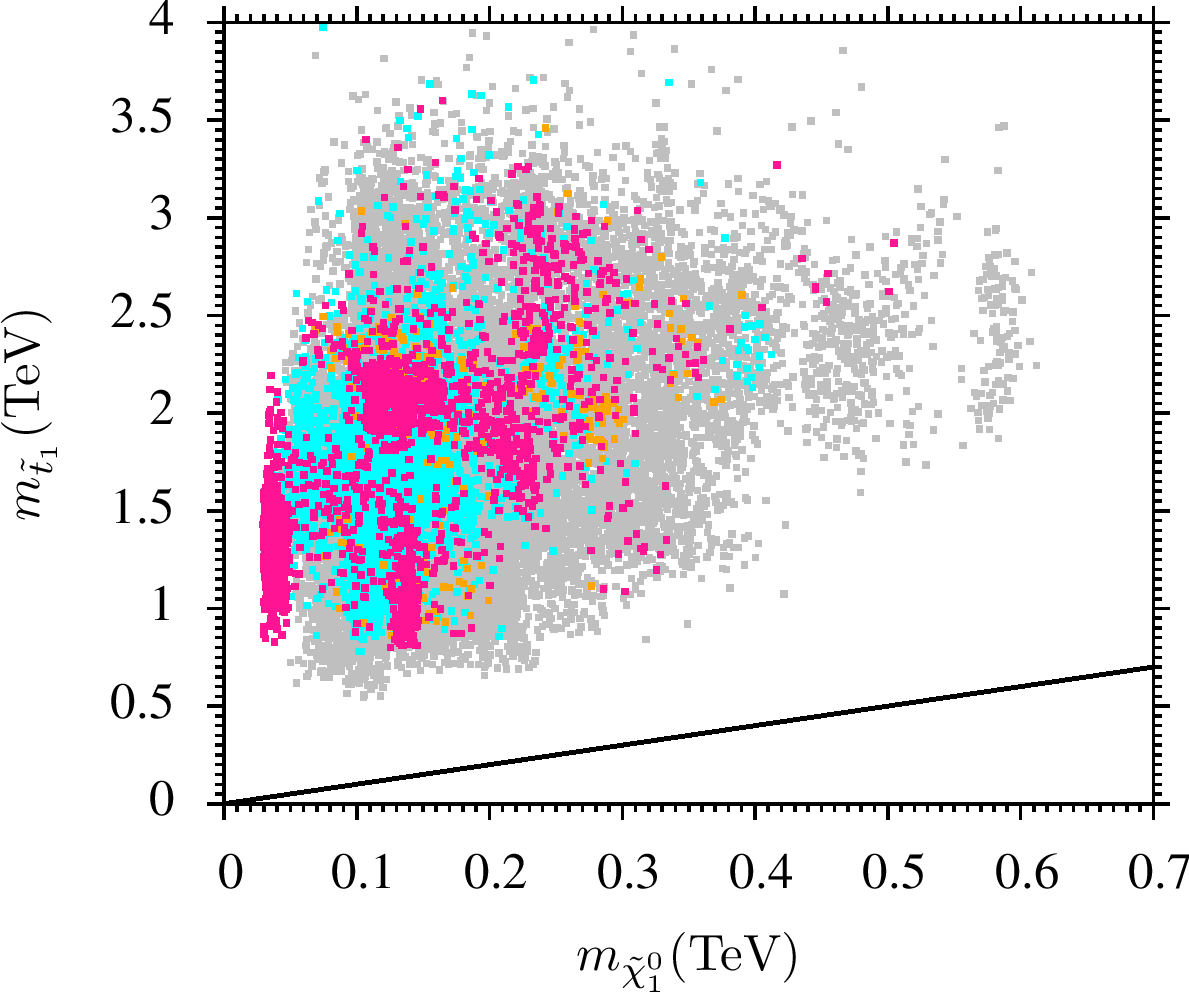}
}
\caption{$m_A$, $m_{\tilde{\chi}_1^{\pm}}$, $m_{\tilde{\tau}_1}$, and $m_{\tilde{\nu}_{\tau}}$ vs. $m_{\tilde{\chi}_1^0}$: Color coding is the same as Figure \ref{fig:tanBetaVSmA}.
}
\label{fig:mAVSmN1}
\end{figure}

Figure \ref{fig:mAVSmN1} display plots in the $m_A-m_{\tilde{\chi}_1^0}$, $m_{\tilde{\chi}_1^{\pm}}-m_{\tilde{\chi}_1^0}$, 
$m_{\tilde{\tau}_1}-m_{\tilde{\chi}_1^0}$ and $m_{\tilde{t}_1}-m_{\tilde{\chi}_1^0}$ plane. Color coding is the same as Figure \ref{fig:tanBetaVSmA}. The black lines in these plots are just to guide the eyes of the reader for the scenario where the lightest neutralino is equal to 
, $\tilde \chi_{1}^{\pm}$, $\tilde \tau_1$, or $\tilde t_{1}$ except in $m_A-m_{\tilde{\chi}_1^0}$ plane where the black line represents 
$m_A= 2 m_{\tilde \chi_{1}^0}$ that is A-resonance condition. Though we do not apply relic density bounds such as reported by WMAP~\cite{Hinshaw:2012aka}, it is expected that some points along the line representing annihilation or coannihilation scenario of the lightest neutralino 
may have relic density within acceptable range.

In the top left panel, we can see how the Higgs coupling precision measurements at HL-LHC (cyan points), ILC (orange points) and CEPC (2IP) (red points) can constrain $m_A$ as has been noted earlier. We also notice that without Higgs coupling constraint,
the lightest neutralino can be as heavy as 0.6 TeV but the CEPC (2IP) constraints restrict it within 0.5 TeV consistent with the observation in Figure~\ref{fig:AmuVSer}. It is also visible that for the $A$-resonance scenario we need $m_{\tilde \chi_{1}^0}$ in the mass range of $\sim [0.3,0.4]$ TeV, 
which also implies $m_A$ should be around $2 m_{\tilde \chi_{1}^0}$ that is $\sim [0.6,0.8]$ TeV. The Higgs coupling precision measurement at ILC can exclude this scenario. We have shown one benchmark point as an example of $A$-resonance solution with correct
relic density as Point 3 in Table \ref{table2a}. 

In the top right panel, we see that points satisfying the constraints mentioned above are overlapped and there is no preferred
parameter region for a given constraint. We also note that all the points along line in this figure
are either Wino-type or Higgsino-type neutralinos with small relic density. It is worth noting our scans show that for the points along the line, most of them are Higgsino-type LSP in the range from 100 GeV to 450 GeV, and only a few points are Wino-type LSP with small mass from 100 to 200 GeV.

In the bottom left panel, we see that grey points have 
$m_{\tilde \tau_{1}}$ mass range anywhere between $[0.14,1.4]$ TeV. After applying the Higgs coupling constraints, $m_{\tilde \tau_{1}}$
confines within 1 TeV. Points along the line represents neutralino-stau coannihilation scenario and some of the points do have correct
relic density but others have low density.

In the bottom right plot, we see that by the application of constraint, $m_{\tilde t_1}$ lies
in the range of $[0.8,4]$ TeV. We have noticed that in our present scans, when $m_{\tilde t_1} \leq$ 2 TeV, $\tilde t_1$ 
is the lightest colored sparticle and when $m_{\tilde t_1} \geq$ 2 TeV, $\tilde g$ is the lightest colored sparticle.
We also notice that there are large mass gaps between $m_{\tilde t_1}$ and $m_{\tilde \chi_{1}^0}$, which suggest that 
$\tilde t_{1,2} \rightarrow t\tilde g$ or $\tilde t_{1,2} \rightarrow t \tilde \chi_{i}^{0}$  or $\tilde t_{1,2} \rightarrow b \tilde 
\chi_{j}^{\pm}$ for the collider searches of stops. Here we want to comment on $\tilde t_{1,2} \rightarrow t \tilde \chi_{i}^{0}$ channel, for this boosted top scenario, 
future colliders like CEPC-SPPC \cite{Arkani-Hamed:2015vfh}, can discovered (excluded) stops up to $\sim$ 6 (8) TeV. This means that all of our points
shown in this plots can be probed at the future colliders. This is one of the examples which shows that the construction of future collider is 
of the utmost need.

\begin{table}[htbp]\hspace{-1.0cm}
\centering
\begin{tabular}{|c|cccc|}
\hline
\hline
                 & Point 1 & Point 2 & Point 3 & Point 4 \\

\hline
$m_{{0}^U}$          &1037    & 2582     &5057   & 5327              \\
$m_{\tilde Q}$      &1003.3  & 2366.4   &4631   & 4879.2 \\
$m_{\tilde U^c}$    &1162.1  & 3306.7   &6487   & 6830.6 \\
$m_{\tilde D^c}$    &1311.7  & 3326.3   &6474.9 & 6841.3   \\
$m_{\tilde L}$      &328.1   & 264.8    &1023   & 858.2    \\
$m_{\tilde E^c}$    &814.1   & 514.8    &900.9  & 978.6      \\
$M_{1} $            &857.1   & 350.9    &602.7  & 686.5   \\
$M_{2}$             &656     & 993      &634.3  & 828.8  \\
$M_{3}$             &1158.8  & -612.25  &555.3  & 473.05 \\
$A_t=A_b$           &-3292   & 5390     &-9095  & -9684   \\
$A_{\tilde \tau}$   &-691.6  & 1192     &-1570  & -2937  \\
$\tan\beta$         &11.6    & 33.8     &19.7   & 21.4    \\
$\mu$               &183     & 172.5    &377.5  & 168.2   \\
$m_{A}$             &338.9   & 754.9    &603    & 846  \\
\hline
$\Delta_{HS}$       &1883    & 3151    & 10873  & 11993.67 \\
$\Delta_{EW}$       &27      & 23      & 87     & 87 \\
$\Delta a_{\mu}$    &$4.731 \times 10^{-10}$ &$16.63\times 10^{-10}$ &$7.683\times 10^{-10}$  &$12.161\times 10^{-10}$  \\
\hline

\hline
$m_h$            &123  & 123 & 125  & 125   \\
$m_H$            &342  &751 & 607  & 856  \\
$m_{H^{\pm}}$    &348  &750 & 608  & 850  \\
\hline
$\kappa_{b},\kappa_{t}$      &1.22541, 0.99817   &1.04295, 0.99996   &1.06459, 0.99983        &1.03296, 0.99992 \\
$\kappa_{\tau}$, $\kappa_{W}=\kappa_{Z}$   &1.22756, 0.99981   &1.04787, 0.99999       & 1.06644, 0.99999    &1.03454, 0.99999 \\
$\kappa_{g},\kappa_{\gamma}$                &0.99083, 1.00369   &1.00473, 0.99763    &0.99954, 1.00008    &1.00038, 0.99714   \\
\hline
$m_{\tilde{\chi}^0_{1,2}}$
                 &171, 191   &131, 183   &260, 376      &161, 178    \\

$m_{\tilde{\chi}^0_{3,4}}$
                 &368, 541   &195, 834    &390, 567  &312, 717    \\

$m_{\tilde{\chi}^{\pm}_{1,2}}$
                 &185, 534   & 179, 823 &378, 555    &175, 702  \\
\hline
$m_{\tilde{g}}$  &2548       &1539  & 1498  & 1317 \\
\hline
$m_{ \tilde{u}_{L,R}}$
                 &2419, 2558   &2705, 3476  & 4721, 6513 &4942, 6832   \\
$m_{\tilde{t}_{1,2}}$
                 &1036, 1798  &1151, 1762     &2311, 3401   &2326, 3543    \\
\hline
$m_{ \tilde{d}_{L,R}}$
                 &2421, 2532   &2706, 3547    &4722, 6595  &4943, 6938   \\
$m_{\tilde{b}_{1,2}}$
                 &1771, 2468    &1212, 3089     &2380, 6340 &2401, 6630    \\
\hline
$m_{\tilde{\nu}_{1,2}}$
                 &726  &521  & 698  & 507 \\
$m_{\tilde{\nu}_{3}}$
                 & 718  &343  &645  & 308 \\
\hline
$m_{ \tilde{e}_{L,R}}$
                &737, 525   &514, 787   &651, 1409   &398, 1487   \\
$m_{\tilde{\tau}_{1,2}}$
                & 518, 728    &353, 568    &627, 1338  &185, 1333   \\
\hline



$\Omega_{CDM}h^{2}$&  0.0014 &0.0846  &0.1017  &0.0099   \\
\hline
\hline
\end{tabular}
\caption{Sparticle and Higgs masses are in GeV units and sign($\mu >$) 0. All of these points satisfy the
constraints \textbf{(\textit{\Rmnum{1}})} and \textbf{(\textit{\Rmnum{2}})} described in Section~\ref{sec:scan}.
Point 1 displays an example of solutions with large deviation in $\kappa_{b}$ and $\kappa_{\tau}$.
Point 2 represents a points with large value of $\Delta a_{\mu}$. Point 3 shows a solution
with $m_A$-resonance while Point 4 is an example of parameter space where stau is degenerated in mass with the lightest neutralino.
}
\label{table2a}
\end{table}

A few mass spectra in the typical region of parameter space after applying all constraints are shown in Table~\ref{table2a}. All the masses in the table are in units of GeV. All of these 
points satisfy the constraints (\textit{\Rmnum{1}}) and (\textit{\Rmnum{2}}) described in Section~\ref{sec:scan}.
Point 1 displays an example of solutions with large deviation in $\kappa_{b} \approx$ 1.22541 and $\kappa_{\tau} \approx$ 1.22756 since $m_A$ 
is small $\sim$ 339 GeV while the deviations of other Higgs couplings are not significant. The measure of electroweak fine-tuning $\Delta_{EW}$ is about 27 but high scale fine-tuning measure is as large as
1883 while $\Delta a_{\mu}$ is about $4.731 \times 10^{-10}$. 
We notice that for all the four points, the SM-like Higgs mass $m_{h}$, heavy CP-even Higgs mass $m_{H}$ and charged Higgs mass $m_{H^{\pm}}$ have masses are in the mass range of $[123, 125]$ GeV and $[340, 860]$ GeV, respectively. The lightest neutralino which is higgsino like, is about 171 GeV while $\chi_{1}^{\pm}$ is about 185 GeV. In the colored sector,
we observe that for all four benchmark points, $m_{\tilde u_{L} }\approx m_{\tilde d_{L}}$, $m_{\tilde u_{R} }\approx m_{\tilde d_{R}}$ 
and their masses are $\geq$ 2.3 TeV.
We also note that for Point 1, gluino mass is about 2.5 TeV which is comparable to the first two generation squark masses, but for other three benchmark
points $m_{\tilde g} <  m_{\tilde q}$. Besides, only for Point 1 the mass difference between $\tilde t_{1}$ and $\tilde b_{1}$ is somewhat large $\sim$ 700 GeV.

It can also be noticed that for Point 1 and Point 2 the light stop is the lightest colored sparticle while for Point 3 and Point 4 gluino is the lightest color sparticle. Slepton masses for Point 1 are less than 800 GeV.

Point 2 represents a point with large value of 
$\Delta a_{\mu} \approx 16.63 \times 10^{-10}$, while $\Delta_{EW}$ and $\Delta_{HS}$ are about 23 and 3151 respectively. Since $m_{A}$ is
relatively small as 755 GeV, the deviations
in $\kappa_{b,\tau}$ are about 4$\%$ and can be probed by CEPC. In electroweakino sector, the lightest neutralino which is bino-higgsino mixed
state, is about 131 GeV while the lightest chargino is about 180 GeV. The gluino and light stop are about 1.5 TeV and 1.1 TeV, respectively.
Since sneutrinos $\tilde \nu_{1,2}$ and smuons $m_{\tilde \mu_{L,R}}$ are as light as $\approx$ 521 GeV and 514 GeV respectively, we observe large $\Delta a_{\mu}$
as noted above. The third generation sneutrino and light stau are about 343 GeV and 353 GeV. Since it is a bino-higgsino mixed point,
its relic density (0.0846) is better than that in Point 1.

Point 3 shows a solution with $m_A$-resonance with $m_{A}\approx$ 603 GeV and correct
relic density $\Omega h^{2} \sim$ 0.1017. Deviations
in $k_{b,\tau}$ are about 6$\%$ and within the range of CEPC measurements. $\Delta_{EW} \sim$ 87, but $\Delta_{HS} \sim$ 10873. 
Gluino and light stop masses are 1.4 TeV and 2.4 TeV respectively. Since electroweakinos and sleptons are relatively heavy as compared to Point 2,  $\Delta a_{\mu}$ is also relatively small 
$7.683 \times 10^{-10}$.

Point 4 is an example where stau is degenerate in mass with the lightest neutralino. Since this
is higgsino-like neutralino, relic density is very small 0.0099. The pseudo-scalar mass $m_A \approx$ 846 GeV, so the $k_{b,\tau}$ are
 as small as about 3$\%$ which can be tested by the CEPEC (2IP) measurements (see Table~\ref{tab:HiggsCplP}). Light stop is about 2.4
TeV while gluino is about 1.3 TeV. Since sneutrino and smuon are light, we have relatively large value for $\Delta a_{\mu} \approx 12.161 \times 10^{-10}$.

\section{Conclusion}
\label{sec:summary}

In this paper, we study the parameter space of natural SUSY in the GmSUSGRA model using the Higgs couplings and the muon anomalous magnetic moment measurements as constraints. Restricting the EWFT measure $\Delta_{\rm EW} \leq 100$, we scan the parameter space by applying basic constraints, the muon anomalous magnetic moment constraint and Higgs coupling constraints from Higgs coupling measurements at HL-LHC, ILC, and CEPC. We display our results in
various plots showing that the Higgs coupling precision measurements and the
muon anomalous magnetic moment measurements can constrain EWSUSY in
GmSUGRA effectively. We show $k_b$ and $k_{\tau}$ as a function of $m_A$. The measurements of the Higgs couplings at the future 
colliders like CEPC can constrain $m_A$ up to 1.2 TeV. In addition to it, light stop mass can  be constrained around 0.5 TeV by the $hgg$
coupling measurements at the CEPC. The combination of muon anomalous magnetic moment and Higgs coupling measurements constrain the $\tilde{e}_R$ mass to be in the range between 600 GeV and 2 TeV. The range of both $\tilde{e}_L$ and $\tilde{\nu}_e$ masses is $[0.4, 1.1]$ TeV. In all cases, the $\tilde{\chi}_1^0$ mass needs to be small (mostly $\leq$ 400 GeV). Since the direct search of heavy Higgs usually excludes the parameter space with large $\tan\beta$ where the branching ratio of $A \rightarrow \tau^{+} \tau^{-}$ can be enhanced, but the $m_A$ bounds given by the Higgs coupling measurement can exclude the region with small $\tan\beta$ as well, the Higgs coupling measurement is complementary to the direct search of heavy Higgs when constraining SUSY. We also list four interesting benchmark points as examples of solutions with large deviations in $\kappa_{b}$ and $\kappa_{\tau}$, large value of $\Delta a_{\mu}$, $m_A$-resonance and stau-neutraino coannihilation in Table \ref{table2a}.

\section*{Acknowledgements}
This research was supported in part by the 
Natural Science Foundation of China under grant numbers 11135003, 11275246, and 11475238 (TL).
K.W. is supported in part by the CAS Center for Excellence in Particle Physics (CCEPP)
and wants to thank Cai-Dian L$\rm \ddot{u}$ for his help.


\bibliographystyle{JHEP}
\bibliography{references}

\end{document}